\documentclass[final,twocolumn,3p,times,sort&compress]{elsarticle}
\usepackage[T1]{fontenc}
\usepackage{txfonts} %

\usepackage[utf8]{inputenc}

\usepackage{amssymb}
\usepackage{color}
\usepackage{url}
\usepackage{hyperref}
\usepackage{graphicx,array} \graphicspath{{figures/}}
\usepackage{subcaption}
\usepackage{amsmath}
\usepackage{siunitx}
\usepackage{bm}
\usepackage{aasmacros}
\usepackage{booktabs}
\usepackage{enumitem} %

\usepackage{lineno}

\newcommand{\figref}[1]{Figure~\ref{fig:#1}}
\newcommand{\tabref}[1]{Table~\ref{tab:#1}}
\newcommand{\secref}[1]{Section~\ref{sec:#1}}
\newcommand{\eqnref}[1]{(\ref{eqn:#1})}

\renewcommand{\b}[1]{{\bm{#1}}}   %

\newcommand{\g}[1]{\b{#1}}
\newcommand{\G}{\mathcal{G}}
\newcommand{\V}{\mathcal{V}}
\newcommand{\E}{\mathcal{E}}
\newcommand{\C}{\mathcal{C}}
\newcommand{\B}{\mathcal{B}}
\renewcommand{\L}{\b{L}}
\newcommand{\tL}{\tilde{\L}}
\newcommand{\W}{\b{W}}
\newcommand{\I}{\b{I}}
\newcommand{\D}{\b{D}}
\newcommand{\U}{\b{U}}
\newcommand{\x}{\b{x}}
\newcommand{\X}{\b{X}}
\newcommand{\y}{\b{y}}
\newcommand{\Y}{\b{Y}}

\newcommand{\f}{\b{f}}
\newcommand{\trans}{^\intercal}
\newcommand{\R}{\mathbb{R}}
\newcommand{\bLambda}{\b{\Lambda}}

\newcommand{\bO}{\mathcal{O}}
\newcommand{\T}{\mathcal{T}}
\DeclareMathOperator*{\esp}{E}
\DeclareMathOperator*{\var}{Var}
\DeclareMathOperator*{\vect}{vec}
\DeclareMathOperator*{\argmin}{arg \, min}
\newcommand{\pkg}[1]{\texttt{#1}}

\journal{Astronomy and Computing}

\begin{document}

\begin{frontmatter}

\title{DeepSphere: Efficient spherical Convolutional Neural Network with HEALPix sampling for cosmological applications}

\author[SDSC]{Nathanaël Perraudin}
\author[EPFL]{Michaël Defferrard}
\author[ETHZ]{Tomasz Kacprzak}
\author[ETHZ]{Raphael Sgier}

\address[SDSC]{Swiss Data Science Center (SDSC), Zurich, Switzerland}
\address[EPFL]{Institute of Electrical Engineering, EPFL, Lausanne, Switzerland}
\address[ETHZ]{Institute for Particle Physics and Astrophysics, ETH Zurich, Switzerland}

\begin{abstract}

Convolutional Neural Networks (CNNs) are a cornerstone of the Deep Learning toolbox and have led to many breakthroughs in Artificial Intelligence.
So far, these neural networks (NNs) have mostly been developed for regular Euclidean domains such as those supporting images, audio, or video.
Because of their success, CNN-based methods are becoming increasingly popular in Cosmology.
Cosmological data often comes as spherical maps, which make the use of the traditional CNNs more complicated.
The commonly used pixelization scheme for spherical maps is the Hierarchical Equal Area isoLatitude Pixelisation (HEALPix).
We present a spherical CNN for analysis of full and partial HEALPix maps, which we call DeepSphere.
The spherical CNN is constructed by representing the sphere as a graph.
Graphs are versatile data structures that can represent pairwise relationships between objects or act as a discrete representation of a continuous manifold.
Using the graph-based representation, we define many of the standard CNN operations, such as convolution and pooling.
With filters restricted to being radial, our convolutions are equivariant to rotation on the sphere, and DeepSphere can be made invariant or equivariant to rotation.
This way, DeepSphere is a special case of a graph CNN, tailored to the HEALPix sampling of the sphere.
This approach is computationally more efficient than using spherical harmonics to perform convolutions.
We demonstrate the method on a classification problem of weak lensing mass maps from two cosmological models and compare its performance with that of three baseline classifiers, two based on the power spectrum and pixel density histogram, and a classical 2D CNN.
Our experimental results show that the performance of DeepSphere is always superior or equal to the baselines.
For high noise levels and for data covering only a smaller fraction of the sphere, DeepSphere achieves typically 10\% better classification accuracy than the baselines.
Finally, we show how learned filters can be visualized to introspect the NN.
Code and examples are available at \url{https://github.com/SwissDataScienceCenter/DeepSphere}.

\end{abstract}


\end{frontmatter}

\begin{figure*}
	\centering
	\includegraphics[width=\linewidth]{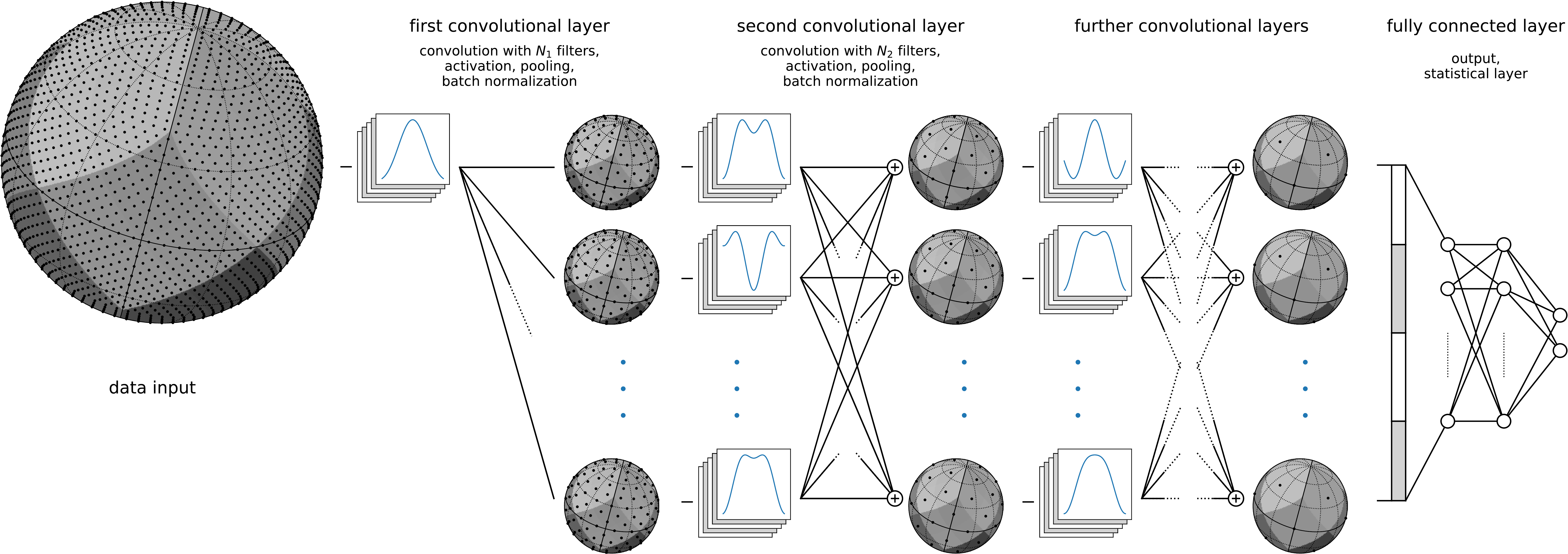}
	\caption{Overall NN architecture, showing here three convolutional layers acting as feature extractors followed by three fully connected layers with softmax acting as the classifier.
    A convolutional layer is based on five operations: convolution, non-linearity, batch normalization, down-sampling, and pooling.
	While most operations are agnostic to the data domain, the convolution and the down-sampling have to be adapted.
	In this paper, we propose first to model the sphere with a graph and to perform the convolution on the graph.
	Graphs are versatile data structures which can model any sampling, even irregular or partial.
	Second, we propose to exploit a hierarchical pixelization of the sphere for the down-sampling operation.
	It allows the NN to analyze the data at multiple scales while preserving the spatial localization of features.
	This figure shows a network that operates on the whole sphere.
	The process is the same when working with partial observations, except that the graph is only built for the region of interest.}
	\label{fig:architecture}
\end{figure*}

\section{Introduction}
\label{sec:intro}

Cosmological and astrophysical data often come in the form of spherical sky maps.
Observables that cover large parts of the sky, such as the Cosmic Microwave Background (CMB) \citep{planck2015cosmologicalparameters,komatsu2011sevenyear,staggs2018recentdiscoveries}, neutral hydrogen \citep{santos2015cosmologySKA,HI4PI2016fullskyHI}, galaxy clustering \citep{alam2017clusteringgalaxies}, gravitational lensing \citep{troxel2017darkenergy,hildebrandt2017kidscosmological}, and others, have been used to constrain cosmological and astrophysical models.
Cosmological information contained in these maps is usually extracted using summary statistics, such as the power spectra or higher order correlation functions.
Convolutional Neural Networks (CNNs) have been proposed as an alternative analysis tool in cosmology thanks to their ability to automatically design relevant statistics to maximise the precision\footnote{Here, the word ``precision'' is to be understood as the final size of the posterior distribution on the measured parameters (including systematic errors).} of the parameter estimation
\citep{schmelze2017cosmologicalmodel,luciesmith2018machinelearning,gupta2018nongaussianinformation,gillet2018deeplearning,hassan2018reionizationmodels,aragoncalvo2018classyfyinglarge,ciuca2017cnnstring,ravanbakhsh2017estimating}, while maintaining robustness to noise.
This is possible as neural networks (NNs) have the capacity to build rich models and capture complicated non-linear patterns often present in the data.
CNNs are particularly well suited for the analysis of cosmological data as their trainable weights are shared across the domain, i.e., the network does not have to relearn to detect objects or features at every spatial location.

\begin{figure*}
\includegraphics[width=\linewidth]{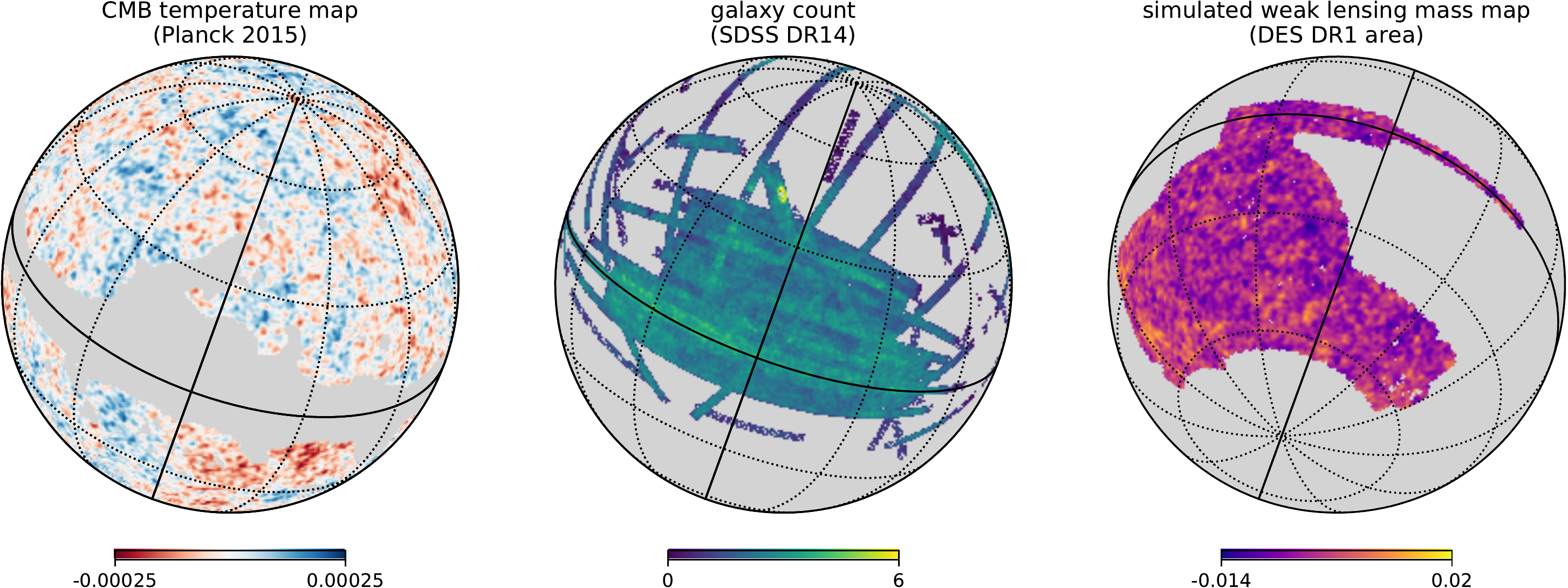}
\caption{Example maps on the sphere:
(left) the CMB temperature (K) map from Planck \citep{planck2015overview}, with galactic plane masked,
(middle) map of galaxy number counts (number of galaxies per arcmin$^2$) in SDSS DR14 \citep{abolfathi2017sdssDR14},
and (right) simulated weak lensing convergence map (dimensionless) simulated with DES DR1 mask \citep{des2018dr1}.
These maps were pixelised using $N_{side} = 512$.
The CMB and weak lensing mass maps were smoothed with Gaussian kernels with FWHM=1 deg, and the galaxy count map with FWHM=0.5 deg.
}
\label{fig:example_maps}
\end{figure*}

So far these algorithms have mostly been demonstrated on Euclidean domains, such as images.
The main challenge in designing a CNN on the sphere is to define a convolution operation that is suitable for this domain, while taking care of the necessary irregular sampling.
Moreover, the designed convolution and resulting NN should possess the following three key characteristics.
First, the convolution should be equivariant to rotation, meaning that a rotation of the input implies the same rotation of the output.
Sky maps are rotation equivariant: rotating a map on the sphere doesn’t change its interpretation.
Depending on the task, we want the CNN to be either equivariant or invariant to rotation.\footnote{When only the statistics of the maps are relevant, they are rotation invariant.}
Second, to be able to train the network in reasonable time, the convolution has to be computationally efficient.
Third, a CNN should work well on parts of the sphere, as many cosmological observations cover only a part of the sky.
For ground-based observations this can be due to limited visibility of the sky from a particular telescope location, and for space-based instruments due to masking of the galactic plane area (see \figref{example_maps} for example maps).

Three ways of generalizing CNNs to spherical data have been pursued.
One approach is to apply a standard 2D CNN to a grid discretisation of the sphere~\citep{boomsma2017spherical,su2017sphericalconv,coors2018spherenet}.
An alternative is to divide the sphere into small chunks and project those on flat 2D surfaces \citep{fluri2018deep,gupta2018nongaussianinformation,schmelze2017cosmologicalmodel,gillet2018deeplearning}.
While these approaches use the well-developed 2D convolution and hierarchical pooling, they are not equivariant to rotation.
Another way is to leverage the spherical Fourier transform and to perform the convolution associated to the SO(3) rotation group in the spectral domain, thanks to the convolution theorem \citep{cohen2018sphericalcnn, esteves2017sphericalcnn}.
While the resulting convolution is equivariant to rotation, this approach is computationally expensive, even if a fast spherical Fourier transform is used.
Moreover, all those methods cannot be much accelerated when maps only span a part of the sky.

Our spherical CNN leverages convolutions on graphs and hierarchical pooling to achieve the following properties: (i) rotation equivariance, (ii) computational efficiency, and (iii) partial sky observations.
The main idea is to model the discretised sphere as a graph of connected pixels: the length of the shortest path between two pixels is an approximation of the geodesic distance between them.
We use the graph CNN formulation introduced in \citep{defferrard2016convolutional}, and a pooling strategy that exploits a hierarchical pixelisation of the sphere to analyse the data at multiple scales.
As the Equal Area isoLatitude Pixelisation (HEALPix)~\citep{gorski2005healpix} is a popular sampling used in cosmology and astrophysics, we tailored the method to that particular sampling.
DeepSphere is, however, easily used with other samplings as only two elements depend on it: (i) the choice of neighboring pixels when building the graph, and (ii) the choice of parent pixels when building the hierarchy.
The flexibility of modeling the data domain with a graph allows one to easily model data that spans only a part of the sphere, or data that is not uniformly sampled.
Using a $k$-nearest neighbours graph, the convolution operation costs $\mathcal{O}(N_{pix})$ operations, where $N_{pix}$ is the number of pixels.
This is the lowest possible complexity for a convolution without approximations.
DeepSphere is readily apt to solve four tasks: (i) global classification (i.e., predict a class from a map), (ii) global regression (i.e., predict a set of parameters from a map), (iii) dense classification (i.e., predict a class for each pixel of a map), and (iv) dense regression, (i.e., predict a set of maps from a map).
Input data are spherical maps with a single value per pixel, such as the CMB temperature, or multiple values per pixel, such as surveys at multiple radio frequencies.

We give a practical demonstration of DeepSphere on cosmological model discrimination using maps of projected mass distribution on the sky \citep{chang2017curvedsky}.
These kind of maps can be created using the gravitational lensing technique \citep[see][for review]{BartelmannSchneider2001weak}.
Our maps are similar to the ones used by \citep{schmelze2017cosmologicalmodel}.
In a simplified scenario, we classify partial sky convergence maps into two cosmological models.
These models were designed to have very similar angular power spectrum.
We compare the performance of DeepSphere to three baselines: a 2D CNN, and an SVM classifier that takes pixel histograms or power spectral densities (PSDs) of these maps as input.
The comparison is made as a function of the additive noise level and the area of the sphere used in the analysis.
Results show that our model is always better at discriminating the maps, especially in the presence of noise.
DeepSphere is implemented with \pkg{TensorFlow} \citep{abadi2016tensorflow} and is intended to be easy to use out-of-the-box for cosmological applications.
Many plots and co PyGSP \citep{pygsp} for computations and plots.
The Python Graph Signal Processing package (PyGSP) \citep{pygsp} is used to build graphs, compute the Laplacian and Fourier basis, and perform graph convolutions.
Code and examples are available online.{\interfootnotelinepenalty=10000 \footnote{\url{https://github.com/SwissDataScienceCenter/DeepSphere/}}}

\section{Method}
\label{sec:method}

A CNN is composed of the following main building blocks~\citep{lecun1998cnn}:
(i) a convolution,
(ii) a non-linearity, and, optionally,
(iii) a down-sampling operation,
(iv) a pooling operation, and
(v) a normalization.\footnote{Batch normalization has been shown to help training \citep{ioffe2015batchnorm}. We verified this experimentally in our setting as well.}
Our architecture is depicted in~\figref{architecture} and discussed in greater details in \secref{architecture}. As operations (ii) and (v) are point-wise, they do not depend on the data domain. The pooling operation is simply a permutation invariant aggregation function which does not need to be adapted either. The convolution and down-sampling operations, however, need to be generalized from Euclidean domains to the sphere.

On regular Euclidean domains, such as 1-dimensional time series or 2-dimensional images, a convolution can be efficiently implemented by sliding a localized convolution kernel (for example a patch of $5 \times 5$ pixels) in the signal domain.
Because of the irregular sampling, there is no straightforward way to define a convolution on the sphere directly in the pixel domain; convolutions are most often performed using spherical harmonics.
In our method we also use the spectral domain to define the convolution.
The implementation, however, does not need a direct access to the spectrum, which is computationally more efficient (see \secref{efficient_convolution}).

Down-sampling is achieved on regular Euclidean domains by keeping one pixel every $n$ pixels in every dimension.
That is again not a suitable strategy on the sphere because of the irregular sampling.

The gist of our method is to define the convolution operation on a sphere using a graph, and the down-sampling operation using a hierarchical pixelisation of the sphere.

\subsection{HEALPix sampling}
\label{sec:healpix}

Before doing any numerical analysis on the sphere, one first has to choose a tessellation, i.e., an exhaustive partition of the sphere into finite area elements, where the data under study is quantized.
The simplicity of the spherical form belies the intricacy of global analysis on the sphere: there is no known point set that achieves the analogue of uniform sampling in Euclidean space.
While our method is applicable to any pixelisation of the sphere, two details depend on the chosen sampling: (i) the choice of neighbours in the construction of the graph, and (ii) the choice of parent vertices when coarsening the graph.
As HEALPix~\citep{gorski2005healpix} is our target application, we tailor the method to that particular sampling in the subsequent exposition.
\figref{example_maps} shows three examples of HEALPix maps: the Cosmic Microwave Background \citep{planck2015overview}, galaxies found in Sloan Digital Sky Survey Data Release 14 \citep{abolfathi2017sdssDR14}, and an example simulated mass map on the footprint of Dark Energy Survey Data Release 1 \citep{des2018dr1}.

HEALPix is a particular case of a more general class of schemes based on a hierarchical subdivision of a base polyhedron.
Another example is the geodesic grids which are based on geodesic polyhedrons, i.e., polyhedrons made of triangular faces. A counter-example is the equirectangular projection, which is not constructed from a base polyhedron, although it can be subdivided.
In the particular HEALPix case, the base is a rhombic dodecahedron, i.e., a polyhedron made from 12 congruent rhombic faces.
See \figref{pooling} for an illustration of the base rhombic dodecahedron and its subdivisions.

The HEALPix pixelisation produces a hierarchical subdivision of a spherical surface where each pixel covers the same surface area as every other pixel.
A hierarchy is desired for the data locality in the computer memory.
Equal area is advantageous because white noise generated by the signal receiver gets integrated exactly into white noise in the pixel space.
Isolatitude is essential for the implementation of a fast spherical transform.
HEALPix is the sole pixelisation scheme which satisfies those three properties.

The lowest possible resolution is given by the base partitioning of the surface into $N_{pix} = 12$ equal-sized pixels (right-most sphere in \figref{pooling}).
The resolution changes as $N_{pix} = 12 N^2_{side}$ such that $N_{pix} = 48$ for $N_{side} = 2$ and $N_{pix} = 192$ for $N_{side} = 3$.
High-resolutions maps easily reach millions of pixels.

\subsection{Graph construction}

Our graph is constructed as an approximation of the sphere $S^2$, a 2D manifold embedded in $\mathbb{R}^3$.
Indeed, \citep{belkin2007convergence} showed that the graph Laplacian converges to the Laplace-Beltrami when the number of pixels goes to infinity providing uniform sampling of the manifold and a fully connected graph built with exponentially decaying weights.
While our construction does not exactly respect their setting (the sampling is deterministic and the graph is not fully connected), we empirically observe a strong correspondence between the eigenmodes of both Laplacians (see \ref{sec:comparison_spherical_harmonics}).

From the HEALPix pixelization, we build a weighted undirected graph $\G = (\V, \E, \W)$, where $\V$ is the set of $N_{pix} = |\V|$ vertices, $\E$ is the set of edges, and $\W$ is the weighted adjacency matrix.
In our graph, each pixel $i$ is represented by a vertex (also called vertex) $v_i \in \V$.
Each vertex $v_i$ is then connected to the $8$ (or $7$)\footnote{\label{neighbors}The $12 \times 4 = 48$ pixels at the corner of each rhombus of the base dodecahedron only have 7 neighboring pixels. See Figures~\ref{fig:healpix_graph_4} and~\ref{fig:pooling}.} vertices $v_j$ which represent the neighboring pixels $j$ of pixel $i$, forming edges $(v_i, v_j) \in \E$. Given those edges, we define the weighted adjacency matrix $\W \in \R^{N_{pix} \times N_{pix}}$ as
\begin{equation*}
	\W_{ij} = \begin{cases}
		\exp \left( -\frac{\|\x_i-\x_j\|_2^2}{\rho^2} \right) & \text{if pixels $i$ and $j$ are neighbors,} \\
		0 & \text{otherwise,} \\
	\end{cases}
\end{equation*}
where $\x_i$ is a vector encoding the 3-dimensional coordinates of pixel $i$, and
\begin{equation*}
	\rho = \frac{1}{|\E|} \sum_{(v_i, v_j) \in \E} \|\x_i-\x_j\|_2
\end{equation*}
is the average Euclidean distance over all connected pixels. This weighting scheme is important as distances between pixels are not equal.
Other weighting schemes are possible. For example,~\cite{khasanova2017graphomni} uses the inverse of the distance instead. We found out that the one proposed above works well for our purpose, and did not investigate other approaches, leaving it to future work.
\figref{healpix_graph_4} shows a graph constructed from the HEALPix sampling of a sphere.

\subsection{Graph Fourier basis}

Following~\cite{shuman2013emerging}, the normalized graph Laplacian,
defined as $\L = \I - \D^{-1/2} \W \D^{-1/2}$, is a second order differential operator
that can be used to define a Fourier basis on the graph. Here $\D$ is the diagonal
matrix where $\D_{ii} = \b{d}_i$ and $\b{d}_i = \sum_j \W_{ij}$ is the weighted degree of vertex $v_i$. By construction, the Laplacian is symmetric positive
semi-definite and hence can be decomposed as $\L = \U \bLambda \U\trans$, where $\U = [\b u_1, \ldots, \b u_{N_{pix}}]$ is an
orthonormal matrix of eigenvectors and $\bLambda$ is a diagonal matrix of
eigenvalues. The graph Fourier basis is defined as the Laplacian eigenvectors, motivated by the fact that a Fourier basis should diagonalize the Laplacian operator.
The graph Fourier transform of a signal $\f \in \R^{N_{pix}}$ is simply its projection on the eigenvectors given by
$\hat{\f} = \mathcal{F}_\G \{\f\} = \U\trans \f$. It follows that the inverse graph Fourier transform reads $\mathcal{F}^{-1}_\G \{\hat{\f}\} = \U\hat{\f} = \U \U\trans \f = \f$.
Note that the Fourier modes are ordered in the increasing order of the Laplacian eigenvalues $\bLambda$, which can be interpreted as squared frequencies.
Indeed,
\begin{equation*}
	\bLambda_{ii} = \b u_i\trans \L \b u_i = \sum_{(v_j, v_k) \in \E} \frac{\W_{jk}}{\sqrt{\b d_j \b d_k}} (\U_{ji} - \U_{ki})^2
\end{equation*}
is a measure of the variation of the eigenvector $\b u_i$ on the graph defined by the Laplacian $\L$.

\begin{figure}[t!]
	\centering
	\includegraphics[width=\linewidth]{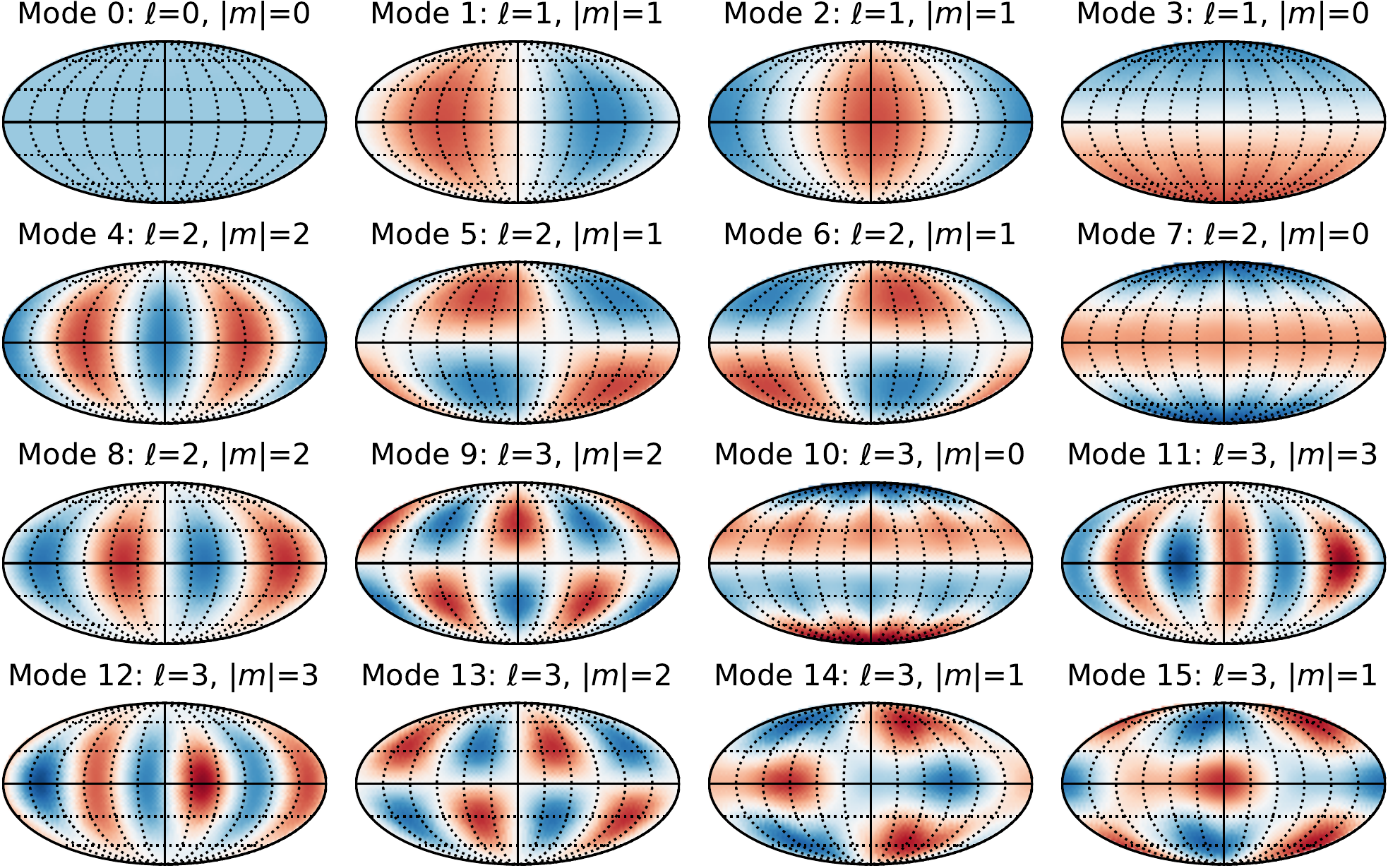}
	\caption{The first 16 eigenvectors of the graph Laplacian, an equivalent of Fourier modes, of a graph constructed from the HEALPix sampling of the sphere ($N_{side}=16$).
    Eigenvectors 1--3 could be associated with spherical harmonics of degree $\ell=1$ and order $|m|=(0,1)$, eigenvectors 4--8 with degree $\ell=2$ and order $|m|=(0,1,2)$, and eigenvectors 9--15 with degree $\ell=3$ and order $|m|=(0,1,2,3)$.
    Nevertheless, graph eigenvectors are only approximating spherical harmonics.}
	\label{fig:graph_harmonics}
\end{figure}

\figref{graph_harmonics} shows the Fourier modes of a HEALPix graph, created using the graph construction described above.
The graph Fourier modes resemble the spherical harmonics.
That is a strong hint that the graph is able to capture the spherical properties of the HEALPix sampling.
This topic is further discussed in \ref{sec:comparison_spherical_harmonics}.

\subsection{Convolution on graphs}
\label{sec:graph_convolution}

As there is no notion of translation on a graph, we cannot convolve two graph signals in a strict sense.
We can, however, convolve a signal with a kernel defined in the spectral domain.
More precisely, we can filter a graph signal by a kernel.
Given the convolution kernel
$h: \R_+ \rightarrow \R$, a signal $\f \in \R^{N_{pix}}$ on the graph is filtered as
\begin{equation} \label{eqn:graph_convolution_fourier}
	h(\L) \f = \U h(\bLambda) \U\trans \f,
\end{equation}
where $h(\bLambda)$ is a diagonal matrix where $h(\bLambda)_{ii} = h(\bLambda_{ii})$.

Contrary to classical signal processing on Euclidean domains, the kernel $h$ has no single representation in the vertex domain and cannot be translated on the graph. It can however be \textit{localized} on any vertex $v_i$ by the convolution with a Kronecker delta\footnote{A Kronecker delta is the signal $\b \delta^i \in \R^{N_{pix}}$ that is zero everywhere except on vertex $v_i$ where it takes the value one.} $\b \delta^i \in \R^{N_{pix}}$. The localization operator $\T_i$ reads $\T_i h = h(\L) \b \delta^i = (h(\L))_i$, the $i$th column of $h(\L)$.
This localization of the kernel $h$ can be useful to visualize kernels, as shown in an example of heat diffusion presented in~\ref{sec:filter_visualization}.
If the graph is not regular, i.e., all vertices do not have the same number of neighbors, and all distances are not equal, the effect of the kernel will slightly differ from one vertex to another. While there is no perfect sampling of the sphere, these differences are negligible as the structure of the whole graph is very regular. However, when considering only parts of the sphere, one can observe important border effects (see ~\ref{sec:border_effects}).

Finally, the graph convolution can be interpreted in the vertex domain as a scalar product with localizations $\T_i h$ of the kernel $h$. Indeed, the result of the convolution of the signal $\f$ with the kernel $h$ is
\begin{equation} \label{eqn:graph_convolution_spatial}
	(h(\L) \f)_i = \langle \T_i h(\L), \f \rangle = \langle h(\L) \b \delta^i, \f \rangle.
\end{equation}
To make the parallel with the classical 1D convolution, let $f, g\in \mathbb{Z} \rightarrow \mathbb{R}$ be two 1D discrete signals. Their convolution can be written in the same form as \eqnref{graph_convolution_spatial}:
\begin{equation*}
	(f \ast g) [i] = \sum_{j=-\infty}^\infty f[j] g[i-j] = \langle T_i g,  f \rangle,
\end{equation*}
where $T_i g[j] = g[i-j]$ is, up to a flip (i.e., $g[i-j]$ instead of $g[i+j]$), a translation operator.
Similarly as \eqnref{graph_convolution_spatial}, the convolution of the signal $f$ by a kernel $g$ is the scalar product of $f$ with translated versions $T_i g$ of the kernel $g$.
Additionally, it turns out that the localization operator $\T_i h$ is a generalization of the translation operator on graphs. In the particular case where the Laplacian matrix $\L$ is circulant, $\T_i h$ is a translated version of $\T_j h$ for all $i, j$ and both convolutions are equivalent. We refer the reader to \citep[Sec 2.2]{perraudin2017stationary} for a detailed discussion of the connection between translation $T_i$ and localization $\T_i$.

To shed some light on the meaning of the convolution on a graph, we show in \ref{sec:heat_diffusion} that the diffusion of heat on a graph can be expressed as the convolution of an initial condition $\f$ with a heat kernel $h$.

\subsection{Efficient convolutions}
\label{sec:efficient_convolution}

\begin{figure}[t!]
    \centering
    \includegraphics[width=\linewidth]{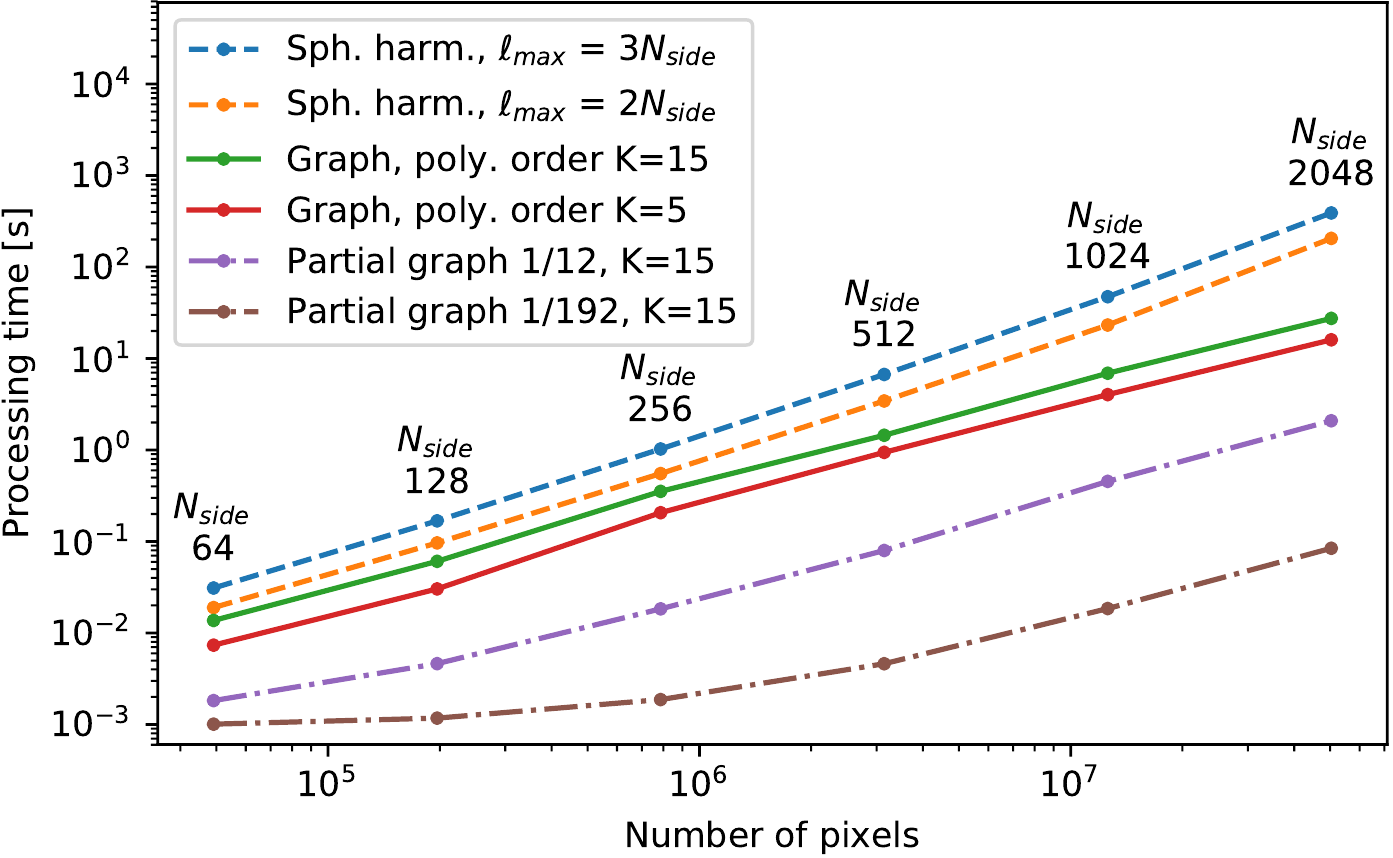}
    \caption{Comparison of filtering speed for Gaussian smoothing of maps of various sizes.
    The fast spherical harmonic transform (SHT) is implemented by the \texttt{healpy} Python package (via the \texttt{healpy.sphtfunc.smoothing} function).
    The graph filtering is defined by \eqnref{graph_convolution_cheby} and implemented with the \texttt{numpy} and \texttt{scipy} Python packages.
    Both are executed on a single core.
	The theoretical cost of filtering on the graph is $\bO(K N_{pix})$ and $\bO(\ell_{max}^3) = \bO(N_{pix}^{3/2})$ for the spherical harmonics, where $\ell_{max}$ is the largest angular frequency.
	The timings for the partial graphs correspond to a convolution on two fractions ($1/12$ and $1/192$) of the sphere, and illustrates the $\bO(N_{pix})$ scaling of graph convolutions.
	}
    \label{fig:filtering_speed}
\end{figure}

While \eqnref{graph_convolution_fourier} is a well justified definition of the convolution, it is computationally cumbersome.
As no efficient and general fast Fourier transform (FFT) exists for graphs \citep{le2018fgft}, the execution of the Fourier transform by multiplication of the signal $\f$ by the dense matrix $\U$ costs $\bO(N_{pix}^2)$ operations.
In a NN, this operation has to be performed for each forward and backward pass.
As current training procedures require processing of many samples, that would be very slow.
Moreover, the eigen-decomposition of the Laplacian $\L$ is needed to obtain the Fourier basis $\U$ in the first place.
That has a unique cost of $\bO(N_{pix}^3)$ operations.

Fortunately, both of these computational issues are overcome by defining the convolution kernel $h$ as a polynomial $h_\theta(\lambda) = \sum_{k=0}^K \theta_k \lambda^k$ of degree $K$ parametrised by $K+1$ coefficients $\theta$.
The filtering operation \eqnref{graph_convolution_fourier} becomes
\begin{equation} \label{eqn:graph_convolution_monomial}
	h_\theta(\L) \f =  \U \left(\sum_{k=0}^K \theta_k \bLambda^k \right) \U\trans \f = \sum_{k=0}^K \theta_k \L^k \f,
\end{equation}
where $\L^k$ captures $k$-neighborhoods.
The entry $(\L^k)_{ij}$ is the sum of all weighted paths of length $k$ between vertices $v_i$ and $v_j$, where the weight of a path is the multiplication of all the edge weights on the path.
Hence, it is non-zero if and only if vertices $v_i$ and $v_j$ are connected by at least one path of length $k$.\footnote{The length of a path between two vertices defines a distance on the graph.}
Filtering with a polynomial convolution kernel can thus be interpreted in the pixel (vertex) domain as a weighted linear combination of neighboring pixel values.
In the classical setting, convolutions are also weighted local sums.
The weights are however given by the filter coefficients only, and there is one coefficient per pixel in the patch.
In the graph setting, the weights are determined by the filter coefficients $\theta$ and the Laplacian $\L$, and there is one coefficient per neighborhood, not vertex.
That defines radial filters, values of which depend only on the distance to the center, and not on the direction.
While it may seem odd to restrict filters to be 1D while the sphere is 2D, radial filters are direction-less and result in rotation equivariant convolutions.
We note that restricting the graph convolutional kernel to a polynomial is similar to restricting the classical Euclidean convolution to a fixed-size patch.
Similarly, each column of the matrix $\sum_{k=0}^K \theta_k \L^k$ defines an irregular patch of radius $K$.
Hence, filters designed as polynomials of the Laplacian have local support in the vertex domain.

Following \citep{defferrard2016convolutional}, we define our filters as Chebyshev polynomials.
The filtering operation \eqnref{graph_convolution_fourier} becomes
\begin{equation} \label{eqn:graph_convolution_cheby}
	h_\theta\left(\tL\right) \f = \U \left(\sum_{k=0}^K \theta_k T_k\left(\tilde{\bLambda}\right) \right) \U\trans \f = \sum_{k=0}^K \theta_k T_k\left(\tL\right) \f,
\end{equation}
where
\begin{equation*}
	\tL = \frac{2}{\lambda_{max}} \L - \b{I} = -\frac{2}{\lambda_{max}} \D^{-1/2} \W \D^{-1/2}
\end{equation*}
is the rescaled Laplacian with eigenvalues $\tilde{\b \Lambda}$ in $[-1, 1]$.
$T_k(\cdot)$ is the Chebyshev polynomial of degree $k$ defined by the recursive relation $T_k(\tL) = 2\tL T_{k-1}(\tL) - T_{k-2}(\tL)$, $T_1(\tL) = \tL$, $T_0(\tL) = \b{I}$.
While definitions \eqnref{graph_convolution_monomial} and \eqnref{graph_convolution_cheby} both allow the representation of the same filters, we found in our experiments that optimizing $\theta$ in \eqnref{graph_convolution_cheby} is slightly more stable than $\theta$ in \eqnref{graph_convolution_monomial}.
We believe that this is due to (i) their almost orthogonality in the spectral and spatial domains, and (ii) their uniformity.\footnote{The amplitude of the Chebyshev polynomials $T_k(x)$ is mostly constant over the domain $[-1, 1]$, independently of the order $k$. On the contrary, the amplitude of the monomials $x^k$ is very different for $|x|\approx 0$ and $|x| \approx 1$.}
Finally, note that while the graph convolution \eqnref{graph_convolution_fourier} is motivated in the spectral domain, definitions \eqnref{graph_convolution_monomial} and \eqnref{graph_convolution_cheby} are implementations in the vertex domain.

Exploiting the recursive formulation of Chebyshev polynomials, evaluating \eqnref{graph_convolution_cheby} requires $\bO(K)$ multiplications of the vector $\f$ with the sparse matrix $\tL$.
The cost of one such multiplication is $\bO(|\E| + |\V|)$.
By construction of our graph, $|\E| < 8 N_{pix}$ and the overall computational cost of the convolution reduces to $\bO(N_{pix})$ operations and as such is much more efficient than filtering with spherical harmonics, even though HEALPix was designed as an iso-latitude sampling that has a fast spherical transform.
This is especially true for smooth kernels which require a low polynomial degree $K$.
\figref{filtering_speed} compares the speed of low-pass filtering for Gaussian smoothing using the spherical harmonics and the graph-based method presented here.
On a single core, a naive implementation of our method is ten to twenty times faster for $N_{side} = 2048$, with $K=20$ and $K=5$, respectively, than using the spherical harmonic transform at $\ell_{max} = 3 N_{side}$ implemented by the highly optimized \pkg{libsharp} \citep{reinecke2013libsharp} library used by HEALPix.
The further important speed-up of graph convolutions on fractions of the sphere is a direct reflection of the $\bO(N_{pix})$ complexity.
On graphs, you only pay for the pixels that you use.
Note that the two might however scale differently, given that \pkg{libsharp} can be distributed on CPUs through MPI while our method can be distributed on GPUs by \pkg{TensorFlow}.

\subsection{Coarsening and Pooling}

Coarsening can be naturally designed for hierarchical pixelisation schemes, where each subdivision divides a cell in an equal number of child sub-cells.
To coarsen, the sub-cells are merged to summarise the data supported on them.
Merging cells lead to a coarser graph.
Coarsening defines $\C(i)$, the set of children of vertex $v_i$.
For the HEALPix subdivision scheme, the number of children is constant, i.e., $| \C(i) | = 4^p \ \forall i$, for some $p$.

\begin{figure}[t!]
	\centering
	\includegraphics[width=\linewidth]{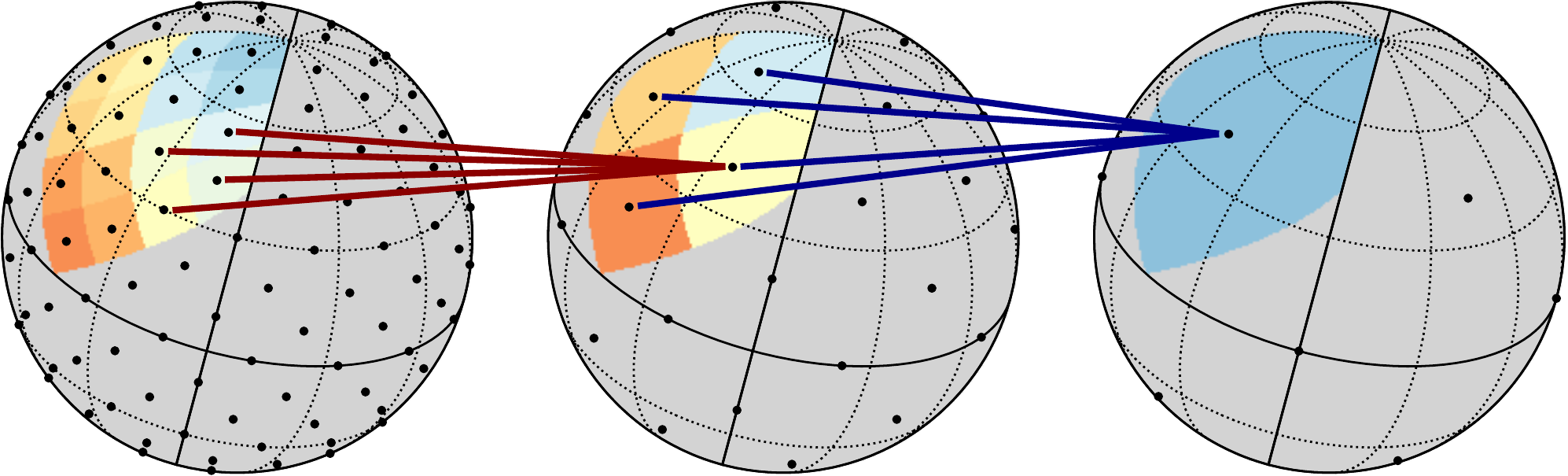}
	\caption{Two levels of coarsening and pooling: groups of 4 cells are merged into one, then the data on them is summarized in one. The coarsest cell covers $1/12$ of the sphere.}
	\label{fig:pooling}
\end{figure}

Pooling refers to the operation that summarizes the data supported on the merged sub-cells in one parent cell.
Given a map $\x \in \R^{N_{pix}}$, pooling defines $\y \in \R^{N'_{pix}}$ such that
\begin{equation} \label{eqn:pooling}
	y_i = f \left( \left\{ x_j : j \in \C(i) \right\} \right), \ \forall i \in [N'_{pix}],
\end{equation}
where $f$ is a function which operates on sets (possibly of varying sizes) and $N_{pix} / {N'_{pix}}$ is the down-sampling factor, which for HEALPix is
\begin{equation*}
| \C(i) | = N_{pix} / {N_{pix}}' = (N_{side} / N'_{side})^2 = 4^p,
\end{equation*}
where $p=\log_2(N_{side} / N'_{side})$.
That operation is often taken to be the maximum value, but it can be any permutation invariant operation, such as a sum or an average.
\figref{pooling} illustrates the process.

\subsection{Layers}

Neural networks are constructed as stacks of layers which sequentially transform the data from its raw representation to some predictions.
The general DeepSphere architecture, pictured in \figref{architecture}, is composed of many layers. The convolutional part, the head of the NN, is composed of graph convolutional layers ($GC$), pooling layers ($P$), and batch normalization layers ($BN$). The tail is composed of multiple fully connected layers ($FC$) followed by an optional softmax layer ($SM$) if the network is used for discrete classification.
A non-linear function $\sigma(\cdot)$ is applied after every linear $GC$ and $FC$ layer, except for the last $FC$ layer.
That operation is point-wise, i.e., $y_{ij} = \sigma(x_{ij})$ and $y_i = \sigma(x_i)$ for matrices $\X, \Y$ and vectors $\x, \y$.
The rectified linear unit (ReLU) $\sigma(\cdot) = \max(\cdot, 0)$ is a common choice, and is the one we adopted in this contribution.

Given a matrix $\X = [\x_1, \ldots, \x_{F_{in}}] \in \R^{N_{pix} \times F_{in}}$, a $GC$ layer computes $\Y = GC(\X) = [\y_1, \ldots, \y_{F_{out}}] \in \R^{N_{pix} \times F_{out}}$, where $N_{pix}$ is the number of pixels (and vertices), $F_{in}$ is the number of input features, and $F_{out}$ is the number of output features.
Using the efficient graph convolution from \eqnref{graph_convolution_cheby}, each output feature map is computed as
\begin{equation*}
	\y_i = \sum_{j=1}^{F_{in}} h_{\theta_{ij}}(\tL) \x_j + b_i \in \R^{N_{pix}}, \ \ \forall i \in [F_{out}].
\end{equation*}
As such, a $GC$ layer is composed of $F_{in} \times F_{out}$ filters, each parameterized by $K$ numbers (see \secref{efficient_convolution}). A bias term $\b b \in \R^{F_{out}}$ is jointly optimized.

Given a matrix $\X \in \R^{N_{pix} \times F}$, a pooling layer computes $\Y = P(\X) \in \R^{N'_{pix} \times F}$ by reducing its spatial resolution ($N'_{pix} < N_{pix}$) according to \eqnref{pooling}.
The batch normalization layer \citep{ioffe2015batchnorm} computes $\Y = BN(\X)$ such as
\begin{equation*}
	\y_i = \gamma_i \frac{\x_i - \esp(\x_i)}{\sqrt{\var(\x_i) + \epsilon}} + \beta_i, \ \forall i \in [F],
\end{equation*}
where $\gamma_{j}$ and $\beta_{j}$ are parameters to be learned and $\epsilon$ is a constant added for numerical stability. The empirical expectation $\esp(\x_i) \in \R$ and variance $\var(\x_i) \in \R$ are taken across training examples and pixels.

The layer $FC: \R^{F_{in}} \rightarrow \R^{F_{out}}$ is defined as
\begin{equation} \label{eqn:fc_layer}
	\y = FC(\x) = \b W \x + \b b ,
\end{equation}
where $\W \in \R^{F_{out} \times F_{in}}$ and $\b b \in \R^{F_{out}}$ are the parameters to be learned.
Note that the output $\Y \in \R^{N_{pix} \times F_{out}}$ of the last $GC$ is vectorized as $\x = \vect(\X) \in \R^{F_{in}}$ before being fed to the first $FC$, where $F_{in} = N_{pix} \times F_{out}$.

The softmax layer is the last layer in a NN engineered for classification.
Given the output $\x \in \R^{N_{classes}}$ of the last $FC$, called the logits in the deep learning literature, the softmax layer outputs $\y = SM(\x)$ such that
\begin{equation*}
	y_i = \frac{\exp(x_i)}{\sum_{j=1}^{N_{classes}} \exp(x_j)} \in [0, 1], \ \forall i \in [N_{classes}],
\end{equation*}
where $N_{classes}$ is the number of classes to discriminate.
Thanks to the softmax, the output $\y \in \R^{N_{classes}}$ of a NN is a discretised conditional distribution for the class given the data and trained parameters.
That is, $y_i$ is the confidence of the network that the input sample belongs to class $i$.
This last layer is actually normalizing $\x$ into $\y$ such that $\| \y \|_1 = \sum_i y_i = 1$.

\subsection{Network architectures}
\label{sec:architecture}

Given a map $\X \in \R^{N_{pix} \times F_{in}}$, a neural network computes $NN_\theta(\X)$, where $NN$ is a composition of the above layers and $\theta$ is the set of all trainable parameters.
The number of input features $F_{in}$ depends on the data. For the CMB radiation temperature, $F_{in} = 1$. For observations in radio frequencies, $F_{in}$ would be equal to the number of surveyed frequencies. $F_{in}$ might also be the number of slices in the radial direction.

\begin{table}
	\centering
	\begin{tabular}{@{ } l r@{ $\in$ }l @{\hspace{0.8em}} r@{ $\in$ }l @{ }}
		\toprule
		 & \multicolumn{2}{c}{Classification} & \multicolumn{2}{c}{Regression} \\
		Global & $NN_\theta(\X)$ & $\R^{N_{classes}}$ & $NN_\theta(\X)$ & $\R^{F_{out}}$ \\
		Dense & $NN_\theta(\X)$ & $\R^{N_{pix} \times N_{classes}}$ & $NN_\theta(\X)$ & $\R^{N_{pix} \times F_{out}}$ \\
		\bottomrule
	\end{tabular}
	\caption{The output's size of the neural network $NN_\theta(\X)$ depends on the task to be solved. $N_{pix}$ is the number of pixels in a HEALPix map, $N_{classes}$ is the number of classes in a classification task, and $F_{out}$ is the number of variables to be predicted in a regression task. The output size of a global task does not depend on the input size, whereas a dense task asks for one prediction per pixel. A classification task asks for discrete (or probabilistic) predictions, whereas a regression task asks for continuous variables.}
	\label{tab:tasks}
\end{table}

DeepSphere can perform dense or global predictions, for regression or classification.
A typical global classification task is to classify maps into cosmological model classes \citep{schmelze2017cosmologicalmodel}. A typical global regression task is to infer the parameters of a cosmological model  from maps \citep{fluri2018deep,gupta2018nongaussianinformation}. Some dense regression tasks are denoising, interpolation of missing values, or inpainting parts of a map \citep{Inoue2008inpainting}. Segmentation and feature detection \citep{Amsel2007detecting} are examples of dense classification tasks.
The size of the output of the NN depends on the task. See \tabref{tasks}.

Fully convolutional networks (FCNs) have been introduced by  \citep{long2015fcn} and are mostly used for the semantic segmentation of images, a dense classification task.
An example FCN for dense regression is
\begin{equation*}
	\Y = NN_\theta(\X) = (GC \circ \sigma \circ BN \circ GC)(\X) \in \R^{N_{pix} \times F_{out}},
\end{equation*}
where $\circ$ denotes composition, i.e., $(f \circ g)(\cdot) = f(g(\cdot))$.
If $P$ layers are used, they have to be inverted via up-sampling in an encoder-decoder architecture.

The set of input pixels which influence the value of an output pixel forms a \textit{receptive field}.
That field is isotropic and local, i.e., it forms a disk centered around the output pixel.
The field's radius is influenced by the polynomial order $K$ of $GC$ layers (setting how far the convolution operation looks around), and the down-sampling factor of $P$ layers, if any.
This radius should be large enough to capture statistics of interest.
For example, a partial sky observation can provide only limited information of cosmological relevance.
On the other hand, looking at the whole sky is often superfluous and waste computations, as most interactions are local.
A data and task dependent trade-off is to be found.
Note here that while global predictions use, by definition, the whole sky, dense predictions are not necessarily local.

By treating each output pixel independently and in parallel, this architecture is a principled way to perform rotation equivariant operations.
Rotation equivariance means that the rotation operation commutes with the NN, i.e., a rotation of the input implies the same rotation of the output.\footnote{Small errors will however appear as $GC$ layers are not exactly equivariant due to the small discrepancy between the graph Fourier modes and the spherical harmonics. See \ref{sec:comparison_spherical_harmonics}.}

Global tasks can be solved by averaging dense predictions \citep{lin2013globalavgpooling, springenberg2014allconv}.
Global averaging is computed by the layer $AV: \R^{N_{pix} \times F} \rightarrow \R^F$.
Doing so assumes the data is locally independent, and form different observations of an unknown process.
Averaging over independent observations is a common way to reduce variance.
An example FCN for global regression is
\begin{equation*}
	NN = AV \circ GC \circ \sigma \circ GC \circ P \circ \sigma \circ BN \circ GC.
\end{equation*}
Here, the FCN predicts parameters for many overlapping parts of the sky in parallel, then average those predictions to get one set of parameters for the whole sky.
Global predictions are made invariant to rotation on the SO(3) group by averaging dense equivariant predictions.
Rotation invariance means that rotating the input does not impact the prediction.

By replacing the average by a $FC$ layer, one gets the standard convolutional neural network (CNN), which is a generalization of the FCN.
Indeed, the $AV$ layer is a $FC$ layer where all the entries of $\W$ in \eqnref{fc_layer} are equal to $1/N_{pix}$ and $\b b = \b 0$.
By dropping rotation invariance, CNNs learn where to put attention on the domain.
That is useful when pixels are not of the same importance, with respect to the task.
For example, on images, the subject of interest is most often around the center of the picture.
Hence, those pixels are more predictive than the ones in the periphery when considering image classification.
On the sphere, location is important when analyzing weather data on the Earth for example, as oceans and mountains play different roles.
An example of such an architecture is
\begin{equation*}
	NN = FC \circ \sigma \circ F \circ P \circ \sigma \circ BN \circ GC.
\end{equation*}

There are in general two ways to deal with symmetries one wants to be invariant to: (i) build them into the architecture, or (ii) augment the dataset such that a (more general) model learns them.
With respect to rotation invariance, the FCN architecture is of the first kind, while the CNN is of the second kind.
For tasks that are rotation invariant, FCNs hence need less training data as the rotation symmetry is backed in the architecture and need not be learned.
This can be seen as intrinsic data augmentation, as a CNN would need to see many rotated versions of the same data to learn the invariance.
Moreover, FCNs can accommodate maps with a varying number of pixels.
Such global summarization as the $AV$ layer is commonly used along graph convolutions to classify graphs of varying sizes (as it is invariant to vertex permutation) \citep{duvenaud2015gcn, li2015gatedgnn}.
As such, CNNs should only be used if rotation invariance is undesired.

All the above architectures can be used for classification (instead of regression) by appending a $SM$ layer.
An example FCN for global classification is therefore
\begin{equation*}
	NN = SM \circ AV \circ GC \circ \sigma \circ GC \circ P \circ \sigma \circ BN \circ GC.
\end{equation*}
Similarly, we emphasize that the use of $FC$ and $AV$ layers is the sole difference between a NN engineered for global or dense prediction.

\subsection{Training}

The cost (or loss) function $C(\Y, \bar \Y) = C(NN_\theta(\X), \bar \Y)$ measures how good the prediction $\Y$ is for sample $\X$, given the ground truth $\bar \Y$. For a classification task, the cost is usually taken to be the cross-entropy
\begin{equation*}
	C(\Y, \bar \Y) = - \sum_{i=1}^{N_{pix}} \sum_{j=1}^{N_{classes}} \bar y_{ij} \log(y_{ij}),
\end{equation*}
where $\bar \Y \in \R^{N_{pix} \times N_{classes}}$ is the ground truth label indicator, i.e., $\bar y_{ij} = 1$ if pixel $i$ of sample $\X$ belongs to class $j$ and is zero otherwise.
For global prediction, we have $N_{pix} = 1$.
For a regression task, a common choice is the mean squared error (MSE)
\begin{equation*}
	C(\Y, \bar \Y) = \frac{\|\Y-\bar{\Y}\|_2^2}{N_{pix} F_{out}} = \frac{1}{N_{pix} F_{out}} \sum_{i=1}^{N_{pix}} \sum_{j=1}^{F_{out}} (y_{ij} - \bar y_{ij})^2,
\end{equation*}
where $\bar{\Y}$ is the desired output. Again, take $N_{pix} = 1$ for global regression. We emphasize that the cost function and the $SM$ layer are the sole differences between a NN engineered for classification or regression.

The goal of training is to find the parameters $\theta$ of the NN that minimize the risk $R(\theta) = \esp \left[ C \left( NN_\theta(\X), \bar \Y \right) \right]$, where $\esp$ is the expectation over the joint distribution $(\X, \bar \Y)$.
In general, that expectation cannot be computed as the data distribution is unknown. We can however minimize an approximation, the empirical risk over the training set $\left\{ \left( \X_i, \bar \Y_i \right) \right\}_{i=1}^{N_{samples}}$:
\begin{equation*}
	\hat{\theta} = \argmin_\theta \sum_{i=1}^{N_{samples}} C \left(NN_\theta(\X_i), \bar \Y_i \right).
\end{equation*}
The optimization is performed by computing an error gradient w.r.t.\ all the parameters by back-propagation and updating them with a form of stochastic gradient descent (SGD):
\begin{equation*}
	\theta \leftarrow \theta - \frac{\eta}{|\B|} \sum_{i \in \B} \frac{\partial C \left( NN_\theta(\X_i), \bar \Y_i \right)}{\partial \theta} ,
\end{equation*}
where $\eta$ is the learning rate, and $\B$ is the set of indices in a mini-batch. Batches are used instead of single samples to gain speed by exploiting the parallelism afforded by modern computing platforms.

\section{Related work}
\label{sec:related_work}

\subsection{2D convolutional neural networks}

\begin{figure}[t!]
    \centering
    \includegraphics[height=7em]{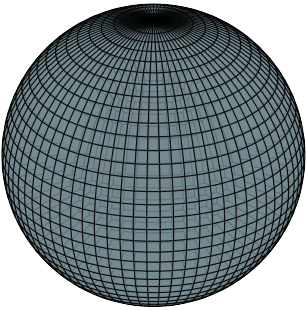}
    \hfill
    \includegraphics[height=7em]{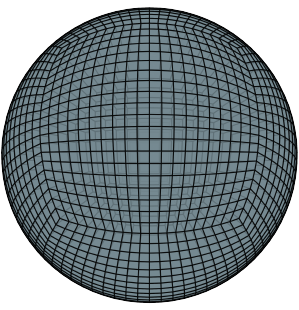}
    \hfill
    \includegraphics[height=7em]{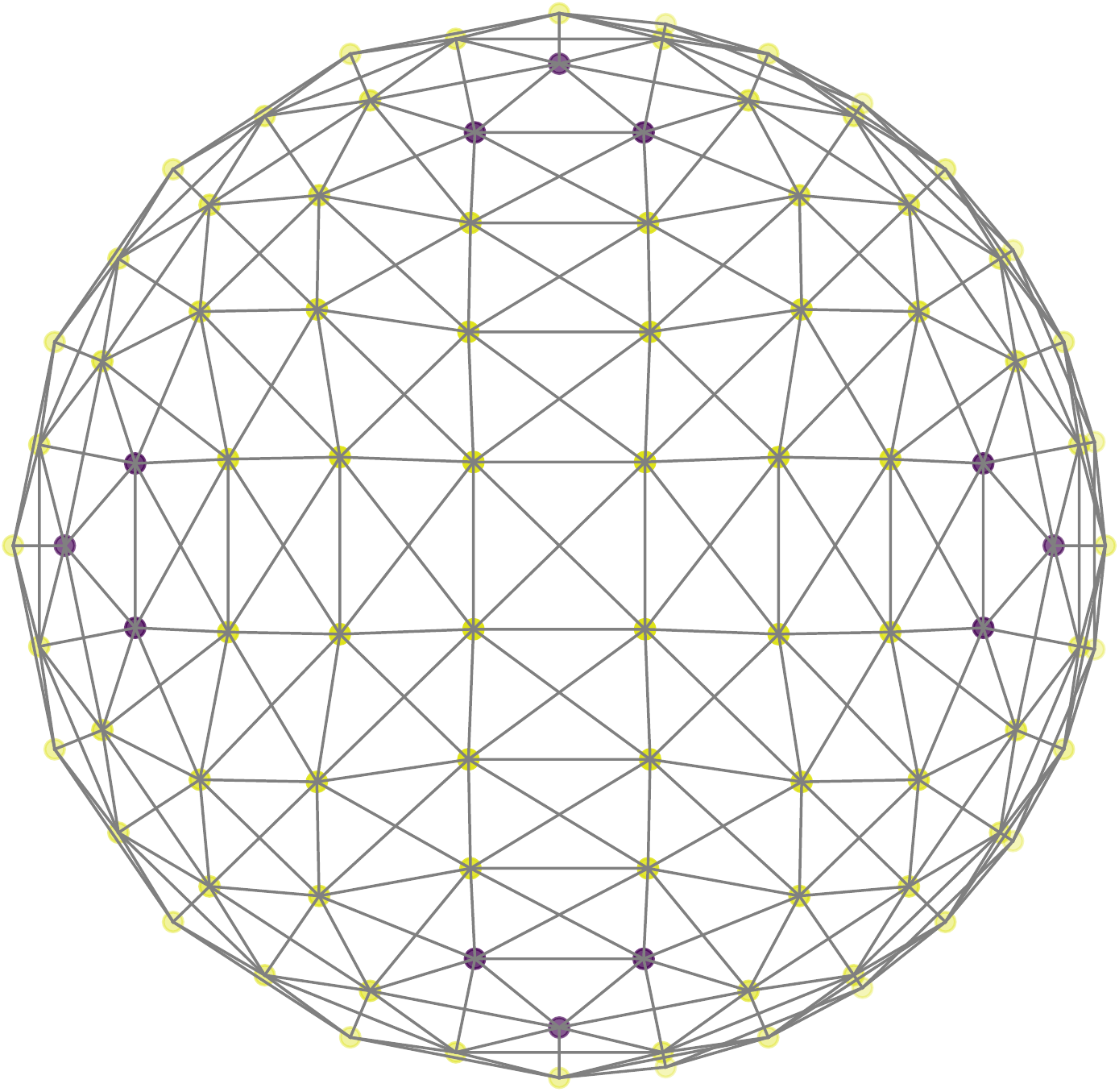}
    \caption[]{Some pixelizations of the sphere. Left: the equirectangular grid, using equiangular spacing in a standard spherical-polar coordinate system. Middle: an equiangular cubed-sphere grid, as described in \citep{ronchi1996cubed}. Right: graph built from a HEALPix pixelization of half the sphere ($N_{side} = 4$). By construction, each vertex has eight neighbors, except the highlighted ones which have only seven.\footnotemark[5] Left and middle figures are taken from \citep{boomsma2017spherical}.}
    \label{fig:sphere_grids}
    \label{fig:healpix_graph_4}
\end{figure}

A first approach, explored by \citep{boomsma2017spherical} for molecular modeling and \citep{su2017sphericalconv, coors2018spherenet} for omnidirectional imaging, is to use a 2D CNN on a discretisation of the sphere that is a grid, such as the equirectangular projection (\figref{sphere_grids}, left panel), or a set of grids, such as the cubed-sphere defined by~\citep{ronchi1996cubed} (\figref{sphere_grids},  middle panel).
As this formulation uses the standard 2D convolution, all the optimizations developed for images can be applied, which makes it computationally efficient.
This approach is applicable to the many pixelizations that are based on a regular subdivision of a base polyhedron, such as HEALPix. Each base polyhedron then forms a grid.
Care has to be taken to handle boundary conditions: for example by padding a grid with the content of the opposite side (equirectangular) or of the neighboring grids (cubed-sphere, HEALPix). That incurs some computational losses.
For samplings that are not equal area, such as the equirectangular projection, the convolution operation should be adjusted to take the induced distortion into account \citep{su2017sphericalconv, coors2018spherenet}.

Another approach to leverage 2D CNNs is to project spherical data onto many tangent planes, which are flat 2D surfaces.
This approach has been extensively used for cosmological maps \citep{fluri2018deep, gupta2018nongaussianinformation, schmelze2017cosmologicalmodel, gillet2018deeplearning} and omnidirectional imaging \citep{xiao2012recognizing, zhang2014panocontext}.
This idea has been generalized to arbitrary 2D manifolds for shape alignment and retrieval \citep{masci2015gcnn, boscaini2016acnn, monti2017monet}.

The main issue with the above two approaches is that they depend on a (local) coordinate system to define anisotropic filters, i.e., filters which are direction dependent.
Direction is well defined and matters for some applications, such as the analysis of weather and climate data on the Earth (north, south, east, west), and omnidirectional imaging (up, down).
Indeed, rotation invariance has been shown to reduce discriminative power for omnidirectional imaging \citep{coors2018spherenet}.
Some problems are, however, intrinsically invariant (or equivariant) to rotation.
Examples include the analysis of cosmological maps, and the modeling of atoms and molecules.
In such cases, directions are arbitrarily defined when setting the origin of the pixelization.
Therefore, a convolution operation has to be isotropic to be equivariant to rotation.

\subsection{Spherical neural networks}

Rotation equivariance was addressed by leveraging the convolution associated to the 3D rotation group SO(3), with applications to atomization energy regression and 3D model classification, alignment and retrieval \citep{cohen2018sphericalcnn,esteves2017sphericalcnn}.
The resulting convolution is performed by (i) a spherical harmonic transform (SHT), i.e., a projection on the spherical harmonics, (ii) a  multiplication in the spectral domain, and (iii) an inverse SHT.
Note the similarity with the naive graph convolution defined in \eqnref{graph_convolution_fourier}.
Likewise, the computational cost of a convolution is dominated by the two SHTs, and a naive implementation of the SHT costs $\bO(N_{pix}^2)$ operations.
Accelerated schemes however exist for some sampling sets \citep[see][for examples]{mohlenkamp1999fast, rokhlin2006fast, reinecke2013libsharp}.
The convolutions remain nevertheless expensive, limiting the practical use of this approach.
For example with HEALPix, which was designed to have a fast SHT by being iso-latitude, the computational cost of the SHT is $\bO(N_{pix}^{3/2}) = \bO(N_{side}^3) = \bO(\ell_{max}^3)$, where $\ell_{max}$ is the largest angular frequency \citep{gorski2005healpix, reinecke2013libsharp}.\footnote{All pixels are placed on $N_{ring} = 4N_{side}-1 = \bO\left(\sqrt{N_{pix}}\right)$ rings of constant latitude.
Each ring has $\bO\left(\sqrt{N_{pix}}\right)$ pixels.
Thanks to this iso-latitude property, the SHT is computed using recurrence relations for Legendre polynomials on co-latitude and fast Fourier transforms (FFTs) on longitude.
The computational cost is thus $\sqrt{N_{pix}}$ FFTs for a total cost of $\bO\left( N_{pix} \log \sqrt{N_{pix}} \right)$, plus $\sqrt{N_{pix}}$ matrix multiplications of size $\sqrt{N_{pix}}$ for a total cost of $\bO\left(N_{pix}^{3/2}\right)$ operations.}
While the Clebsh-Gordan transform can be leveraged to avoid SHTs between layers (it does a non-linear transformation in the spectral domain), the transform itself costs $\bO(N_{pix}^{3/2})$, for no reduction of overall complexity \citep{kondor2018clebsch}.
This work is another indication that there might be a $\Omega(N_{pix}^{3/2})$ computational lower bound for the proper treatment of rotation equivariance with the spherical harmonics.
In comparison, DeepSphere scales as $\bO(N_{pix})$ (see \secref{efficient_convolution}), at the expense of not being exactly equivariant (see \ref{sec:comparison_spherical_harmonics}).
Being a mathematically well-defined rotation equivariant network is the main advantage of methods based on spherical harmonics, though the fact that convolution kernels are learned probably compensates for inexact equivariance.
See \citep{kondor2018equivariance} for a rigorous treatment of convolution and equivariance in NNs.

While defining the convolution in the spectral domain avoids all sampling issues, filters so defined are not naturally localized in the original domain.
Localization is desired for the transformation to be stable to local deformation \citep{mallat2012scattering}, and most interactions in the data are local anyway.
A straightforward way to impose locality in the original domain is to impose smoothness in the spectral domain, by Heisenberg's uncertainty principle.\footnote{Heisenberg's uncertainty principle states that a filter cannot be arbitrarily concentrated in one domain without being de-concentrated in the other.}
A more elegant approach is to define filters that are provably localized.
The filters defined in \eqnref{graph_convolution_monomial} and \eqnref{graph_convolution_cheby}, by being polynomials of the Laplacian, are of this kind.

All the presented 2D and spherical CNNs cannot be easily accelerated when the data lies on a part of the sphere only.
That is an important use case in cosmology as measurements are often partial, i.e., whole sky maps are rare.
One could still fill the unseen part of the sphere with zeros, and avoid computing empty integrals.
It is, however, not straightforward to identify empty space, and computations would still be wasted (for example on a ring that mostly contains zeros but a few measurements).
With graphs, however, computations are only performed for used pixels.
While it results in some distortions due to border effects (see \figref{border_effects} and \ref{sec:border_effects}), these can be mitigated by padding with zeros a small area around the measurements.

\subsection{Graph neural networks}

The use of a graph to model the discretised sphere was also considered for omnidirectional imaging \citep{khasanova2017graphomni}.
This work is the closest to our method, with three differences. First, they parametrize their convolution kernel with \eqnref{graph_convolution_monomial} instead of \eqnref{graph_convolution_cheby} (see \secref{efficient_convolution} for a discussion and comparison).
Second, they did not take advantage of a hierarchical pixelization of the sphere and resorted to dynamic pooling \citep{kalchbrenner2014dcnn}.
While that operation has proved its worth to pool sequences of varying length, such as sentences in language models, it is undesired in our context as it is not local and destroys spatial coherence.
Third, they introduced a statistical layer --- an operation that computes a set of statistics from the last feature maps --- to provide invariance to rotation.
We propose to use the idea of fully convolutional networks (FCNs) instead (see \secref{architecture}).
While statistics have to be hand-chosen to capture relevant information for the task, the filters in a FCN are trained end-to-end to capture it.

Many formulations of graph neural networks, reviewed by \citep{bronstein2017review} and \citep{hamilton2017review}, have been proposed.
For this contribution, we chose the formulation of \citep{defferrard2016convolutional} as its root on strong graph signal processing theory makes the concept of convolutions and filters explicit \citep{shuman2013emerging}.
As the convolution is motivated by a multiplication in the graph Fourier spectrum, it is close in spirit, and empirically, to the formulation based on the spherical harmonics, which is the ideal rotation equivariant formulation.

Thanks to their versatility, graph neural networks have been used in a variety of tasks, such as identifying diseases from brain connectivity networks \citep{ktena2018metriclearning} or population graphs \citep{parisot2017disease}, designing drugs using molecular graphs \citep{hop2018drugdesign}, segmenting 3D point clouds \citep{qi2017pointcloudsegmentation}, optimizing shapes to be aerodynamic \citep{baque2018shape}, and many more.
By combining graph convolutional layers and recurrent layers \citep{seo2016gcrn}, they can, for example, model structured time series such as traffic on road networks \citep{li2018traffic}, or recursively complete matrices for recommendation \citep{monti2017recommendation}.
Another trend, parallel to the modeling of structured data, is the use of graph neural networks for relational reasoning \citep{battaglia2018review}.

\section{Experiments}
\label{sec:experiments}

The performance of DeepSphere is demonstrated on a discrimination problem: the classification of convergence maps into two cosmological model classes.
The experiment presented here is similar to the one by \citep{schmelze2017cosmologicalmodel}.
These maps are similar to those created with gravitational lensing techniques \citep{chang2017curvedsky}.
The two sets of maps were created using the standard cosmological model with two sets of cosmological parameters (see \secref{data} for details).
The classification methods were trained to predict a label from a HEALPix map.
Our Python implementation to reproduce those experiments is openly available online.{\interfootnotelinepenalty=10000 \footnote{\url{https://github.com/SwissDataScienceCenter/DeepSphere}}}
The data is available upon request.\footnote{\url{https://doi.org/10.5281/zenodo.1303272}}

While we only demonstrate a classification task here, other tasks, such as regression or segmentation, are also possible (see \secref{architecture}).
The regression task will most likely be the most practical cosmological application, as described in \citep{gupta2018nongaussianinformation} and \citep{fluri2018deep}.
Implementation of full cosmological inference typically requires, however, many more simulations, building the likelihood function, and several other auxiliary tasks.
The classification problem is much more straightforward to implement and execute, and can be used to fairly compare the accuracy of the algorithm against benchmarks \citep{schmelze2017cosmologicalmodel}.
For these reasons we decided to use the classification problem in this work, and we expect the relative performance of the methods to generalise to the regression scenario.

\subsection{Data}
\label{sec:data}

\begin{figure}[t!]
	\centering
	\includegraphics[width=\linewidth]{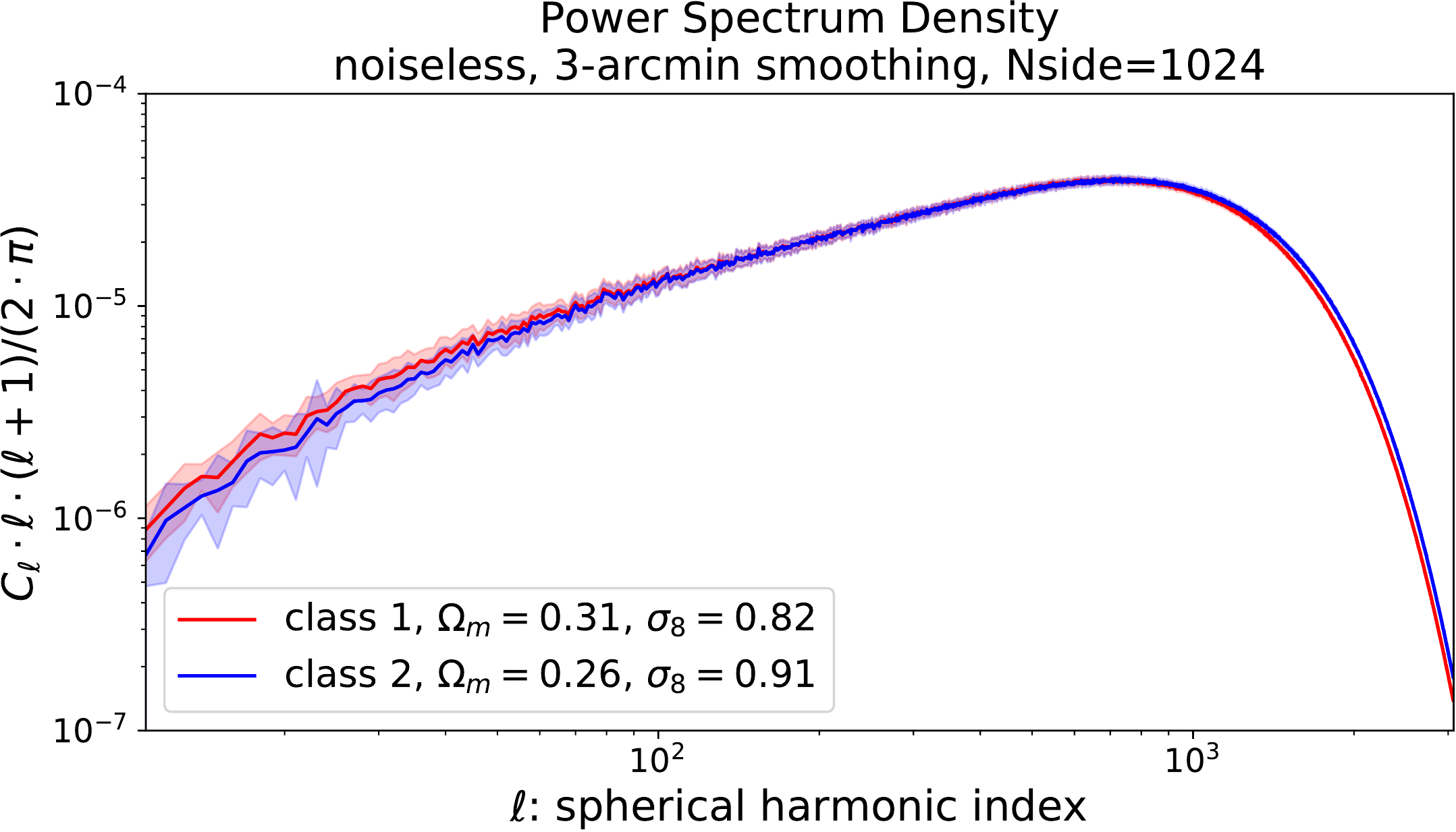}
	\caption{Power spectral densities of the noiseless maps.
		To prevent the cosmological models from being distinguished by their power spectra alone, the maps have been smoothed with a Gaussian kernel of radius $3$ arcmins to remove high frequencies ($\ell>1000$).}
	\label{fig:psd_sigma3}
\end{figure}

\begin{figure*}
	\centering
	\includegraphics[width=0.48\linewidth]{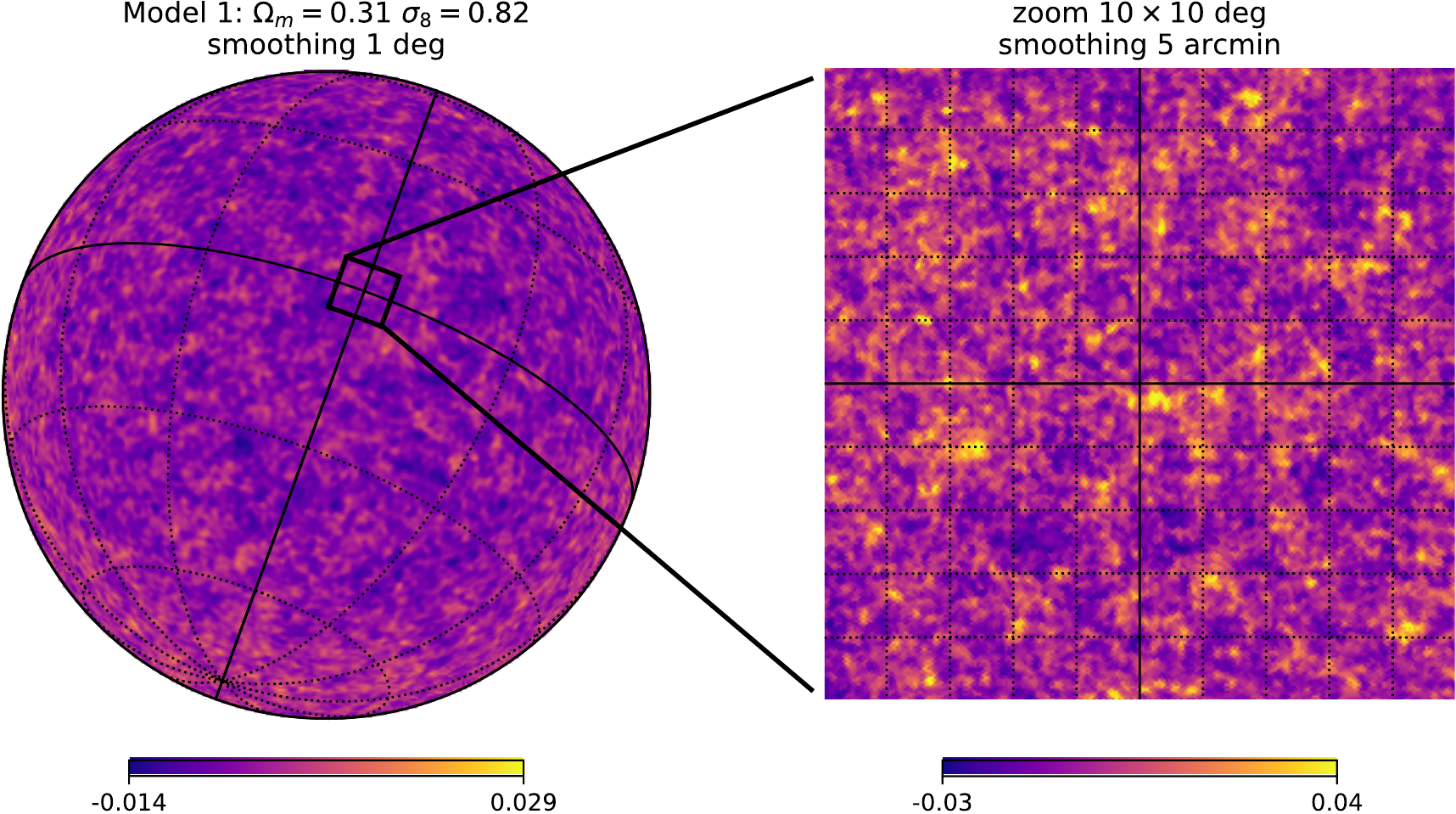}
	\hfill
	\includegraphics[width=0.48\linewidth]{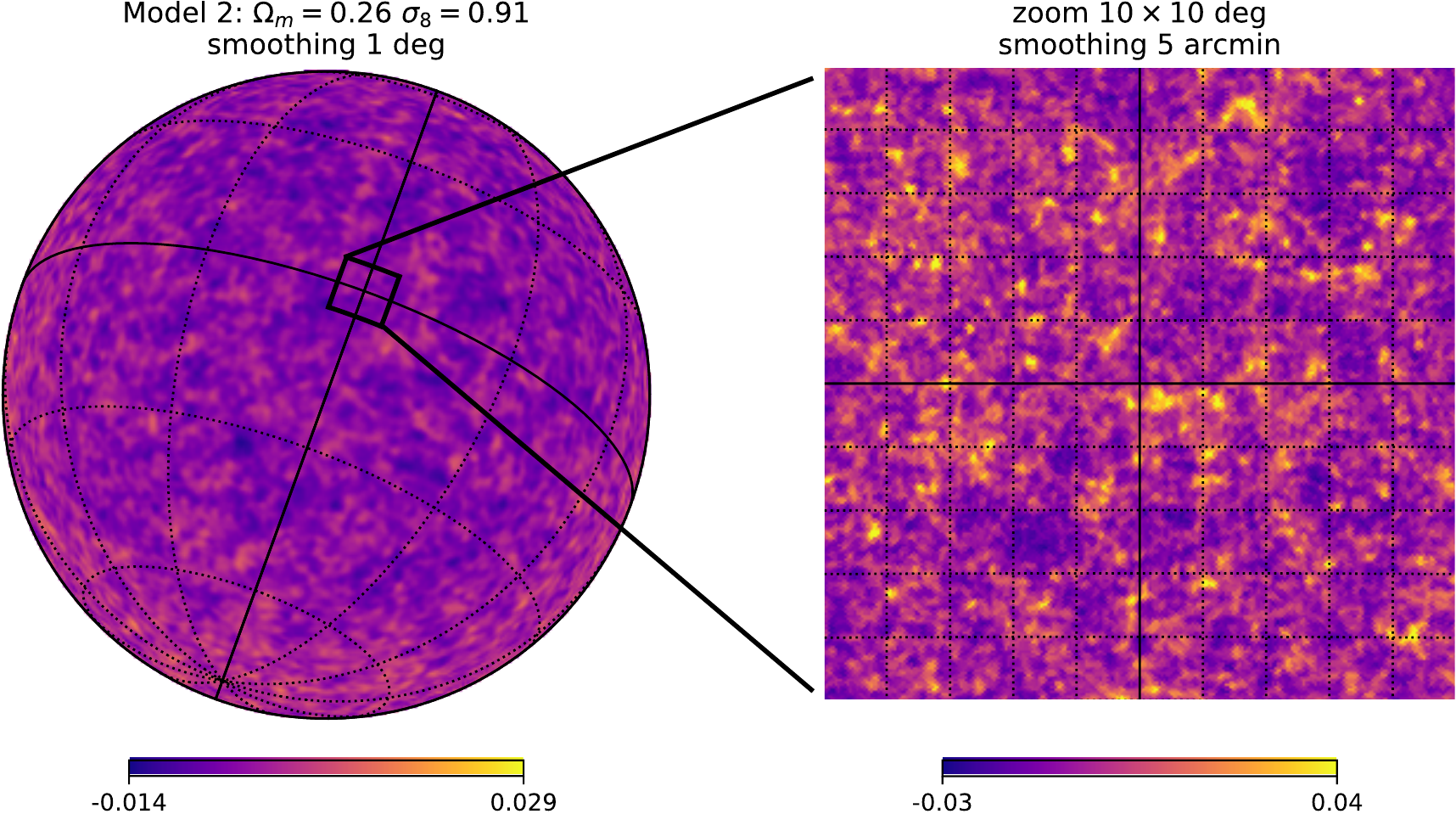}
	\caption{Example maps from two classes to be discriminated. Left: model 1 with $\Omega_m=0.31$ and $\sigma_8=0.82$. Right: model 2 with $\Omega_m=0.26$ and $\sigma_8=0.91$.
	The initial conditions for both simulations are the same, so differences only arise due to different cosmological parameters.}
	\label{fig:map_sample}
\end{figure*}

Convergence maps represent the dimensionless distribution of over- and under-densities of mass in the universe, projected on the sky plane.
The 3D structures are projected using a geometric kernel, the value of which depends on the radial distance.
In gravitational lensing, this kernel is dependent on the radial distances between the observer, the mass plane, and the plane of source galaxies \citep[see][for a review of gravitational lensing]{bartelman2010gravitationallensing}.

We make whole sky N-body simulations for two parameter sets of the $\Lambda\rm{CDM}$ cosmological model: model 1 ($\Omega_m=0.31, \sigma_8=0.82$) and model 2 ($\Omega_m=0.26, \sigma_8=0.91$), where $\Omega_m$ is the matter density in the universe and $\sigma_8$ is the normalisation of the matter power spectrum.
Other parameters are set to: Hubble constant $H_0=70 km/s/Mpc$, spectral index $n_s=0.96$, and Baryon density today $\Omega_b=0.05$.
The parameters $\Omega_m$ and $\sigma_8$ were chosen for the maps to have the same spherical harmonic power spectrum.
That means that it is difficult to distinguish between these cosmological models.
We found that the differences in power spectrum is $~5\%$ for $\ell>1000$.
To remove this information, we additionally smooth the spherical maps with a Gaussian kernel of radius $3$ arcmin.
The resulting power spectral density (PSD), computed using the \texttt{anafast} function of the HEALPix package, are displayed in \figref{psd_sigma3}.
We also subtract the mean of each map and down-sample them to a resolution of $N_{side}=1024$, which corresponds to maps of $12 \times 1024^2 \approx 12 \times 10^6$ pixels.
As shown by the occupied spectrum (\figref{psd_sigma3}), a larger resolution would waste memory and computation, while a lower resolution would destroy information.

The simulations are created using the fast lightcone method \pkg{UFalcon} described in \citep{sgier2018fastgeneration}.
A brief overview about the map making procedure used in \pkg{UFalcon} as well as the simulation parameters are given in \ref{sec:convergence_mass_maps}.
We however use a single simulation box, as opposed to two used in that work, as we use source galaxies at a lower redshift of $z=0.8$, instead of $z=1.5$.
\pkg{L-PICOLA} \citep{howlett2015lpicola} is used for fast and approximate N-body simulations.
We generate $30$ simulations for each of the two classes.
Out of the $60$ simulations, $20$ are kept as the test set, and $20\%$ of the remaining training data is used as a validation set, to monitor the training process and select the hyper-parameters.
\figref{map_sample} shows the whole sky simulations and a zoom region for both models.
Initial conditions for these simulations are the same, so the differences in structures can only be attributed to different cosmological parameters used to evolve the particle distribution.

\subsection{Problem formulation}

As the distribution of matter in the universe is homogeneous and isotropic, no pixel is more informative than any other.\footnote{Contrast that with images, where the subject of interest is most often around the center of the picture.}
As such, we can control the difficulty of the classification problem by limiting the number of pixels available to the algorithms, i.e., extracting partial maps from whole sky maps.
Using the properties of HEALPix, we split maps into $12 \times o^2$ independent samples of $(N_{pix} / o)^2$ pixels (for $o=1,2,4,\dots$). The resulting partial maps are large enough to suffer from the effects of the spherical geometry, and cover 8.3\% ($\approx 1 \times 10^6$ pixels), 2.1\% ($\approx 260 \times 10^3$ pixels), and 0.5\% ($\approx 65 \times 10^3$ pixels) of the sphere for $o=1,2,4$, respectively.
Corresponding areas can be seen in \figref{pooling}: the surface of pixels in the left, middle, and right spheres correspond to samples at order $o=4,2,1$, respectively.
We report results for $o=1,2,4$ only, as full sphere classification is easy at such resolution (perfect classification accuracy is already obtained at $o=1$, see \figref{results}).
The published code nonetheless includes an example demoing classification on full spheres of lower resolution ($N_{side} = 64$).

To make the discrimination harder and the problem more realistic, centered Gaussian noise is added to the simulated maps.
The standard deviation of the noise varies from zero (i.e., no noise) to $2\times$ the standard deviation of pixel's values in the noiseless maps.
While the noise model of real maps often has a slightly different distribution, Gaussian noise should be a sufficient model to demonstrate the performance of our method.
To avoid over-fitting, random noise is generated during training, so that no two samples are exactly the same (i.e., two samples might be from the same simulation, but the added noise will be different).
That is a data augmentation scheme, as it creates more training examples than we have simulations.

\subsection{Baselines}

DeepSphere is first compared against two simple yet powerful benchmarks.
The two baselines are based on features that are (i) the power spectral densities (PSDs) of maps, and (ii) the histogram of pixels in the maps \citep{patton2017cosmologicalconstraints}.
After normalization, those features are used by a linear support vector machine (SVM) trained for classification.
A linear kernel is used as other kernels did not provide better results, while having significantly worse scaling properties.
For fairness, the training set was augmented in a similar way as for DeepSphere: we created samples by adding new random realizations of the noise. We stopped adding new samples to the training data once the validation error converged. This process is detailed in \ref{sec:dataset_augmentation}.
As for DeepSphere, the SVM regularization hyper-parameter was tuned based on the performance over the validation set.
The classification accuracy of DeepSphere and both baselines was checked as a function of the used area (the order $o$), and the relative level of additive noise.

DeepSphere is further compared to a classical CNN for 2D Euclidean grids, later referred to as 2D ConvNet.
To be fed into the 2D ConvNet, samples have to be transformed into flat images.
As described in \ref{sec:projection}, we use the property that HEALPix is defined as the subdivision of 12 square surfaces to efficiently project the spherical maps.
This property has been independently exploited (while this paper was under review) to define another spherical CNN \cite{krachmalnicoff2019convolutional}.
In our experiments, we used the same architecture as DeepSphere (described in \secref{hyper_parameters}), with convolution kernels of size $5 \times 5$ and an $\ell2$ weight decay with a weight of 3 as regularization.
The number of filters was reduced in order to match the number of trainable parameters between DeepSphere and the 2D ConvNet.
(A larger architecture resulted in unstable training for a marginal performance gain.)

Ideally, we would compare DeepSphere to alternative spherical CNNs like \citep{cohen2018sphericalcnn} and \citep{esteves2017sphericalcnn}.
There are however two issues with that comparison.
First, these CNNs were developed for the equirectangular sampling.
An option is to modify the available implementations to use HEALPix, which is a major undertaking due to their complexity.
Another option is to transform the cosmological maps to the equirectangular sampling, which is of disputable usefulness as the field is unlikely to use that sampling.
Second, those methods, based on SHTs, only work on the full sphere (at least in their current state).
As shown in \figref{filtering_speed}, using the full sphere at a resolution of $N_{side} = 1024$ would result in a slow-down of multiple orders of magnitude compared to using a partial graph.
The authors of \citep{he2018cmbdl} could not use \citep{cohen2018sphericalcnn} for CMB analysis because of that.\footnote{Note that reported results in \citep{cohen2018sphericalcnn} and \citep{esteves2017sphericalcnn} are for maps of at most $(2 \times 128)^2 = 65,536$ pixels.
That is the size of our smallest partial map, while the largest is of approximately one million pixels.}
For those reasons and because we think that a proper comparison is better carried out on diverse datasets and tasks, we leave it as future work.

\subsection{Network architecture and hyper-parameters}
\label{sec:hyper_parameters}

As discussed in \secref{architecture}, the choice of a network architecture depends on the data and task at hand.
For our problem, where we assume that each pixel carries the same quantity of information about the cosmological model, a rotation invariant FCN architecture should be best.
The selected network is defined as
\begin{equation*}
	FCN = \underbrace{SM \circ AV \circ GC_2}_{\text{classifier}} \circ \underbrace{L_5 \circ L_4 \circ L_3 \circ L_2 \circ L_1}_{\text{feature extractor}},
\end{equation*}
\begin{center}
with \hspace{0.5cm} \begin{tabular}{lll}
   $L_1$ &  $=$ &$ P_4 \circ \sigma \circ BN \circ GC_{16}$, \\
   $L_2$ &  $=$ &$ P_4 \circ \sigma \circ BN \circ GC_{32}$, \\
   $L_3,L_4,L_5$ & $=$ &$ P_4 \circ \sigma \circ BN \circ GC_{64}$, \\
\end{tabular}
\end{center}
where $GC_F$ indicates a graph convolutional layer with $F$ output feature maps, $P_4$ is a pooling layer that divides the number of pixels by 4, $\sigma$ is a ReLU, and the Chebyshev polynomials in \eqnref{graph_convolution_cheby} are of degree $K=5$.
The layers $L_1$ to $L_5$ build the statistical evidence to classify each pixel of a down-sampled sphere.
The last $GC$ layer acts as a linear classifier and outputs two predictions.
The $AV$ layer averages these predictions and the $SM$ layer normalizes them.

For comparison, we tried the more conventional CNN architecture, where the classification is performed by a fully connected layer from the last feature maps instead of the average of local classifications. This architecture is not invariant to rotation. The selected network is defined as
\begin{equation*}
	CNN = \underbrace{SM \circ FC_2}_{\text{classifier}} \circ \underbrace{L_5 \circ L_4 \circ L_3 \circ L_2 \circ L_1}_{\text{feature extractor}},
\end{equation*}
where $FC_2$ denotes a fully connected layer with two outputs.
Architectures were selected over their performance on the validation set.

The Adam optimizer \citep{kingma2014adam} with $\beta_1=0.9$, $\beta_2=0.999$ was used with an initial learning rate of $2 \times 10^{-4}$ that is exponentially decreased by a multiplication with $0.999$ at each step.
All models were trained for $80$ epochs, which corresponds to $1920$ steps. The batch size is set to $16 \times  o^2$ to keep the amount of supervision used to estimate the gradient identical, irrespective of the chosen order $o$ (the learning rate would otherwise need to be adapted). By this choice, DeepSphere is trained with the same amount of information, i.e., the number of pixels, across all variants of the problem.
Training took approximately $1.5$ hour using a single Nvidia GeForce GTX 1080 Ti and $5$ hours with a Tesla K20.

\subsection{Results}

\begin{figure*}
	\centering
	\includegraphics[width=0.32\linewidth]{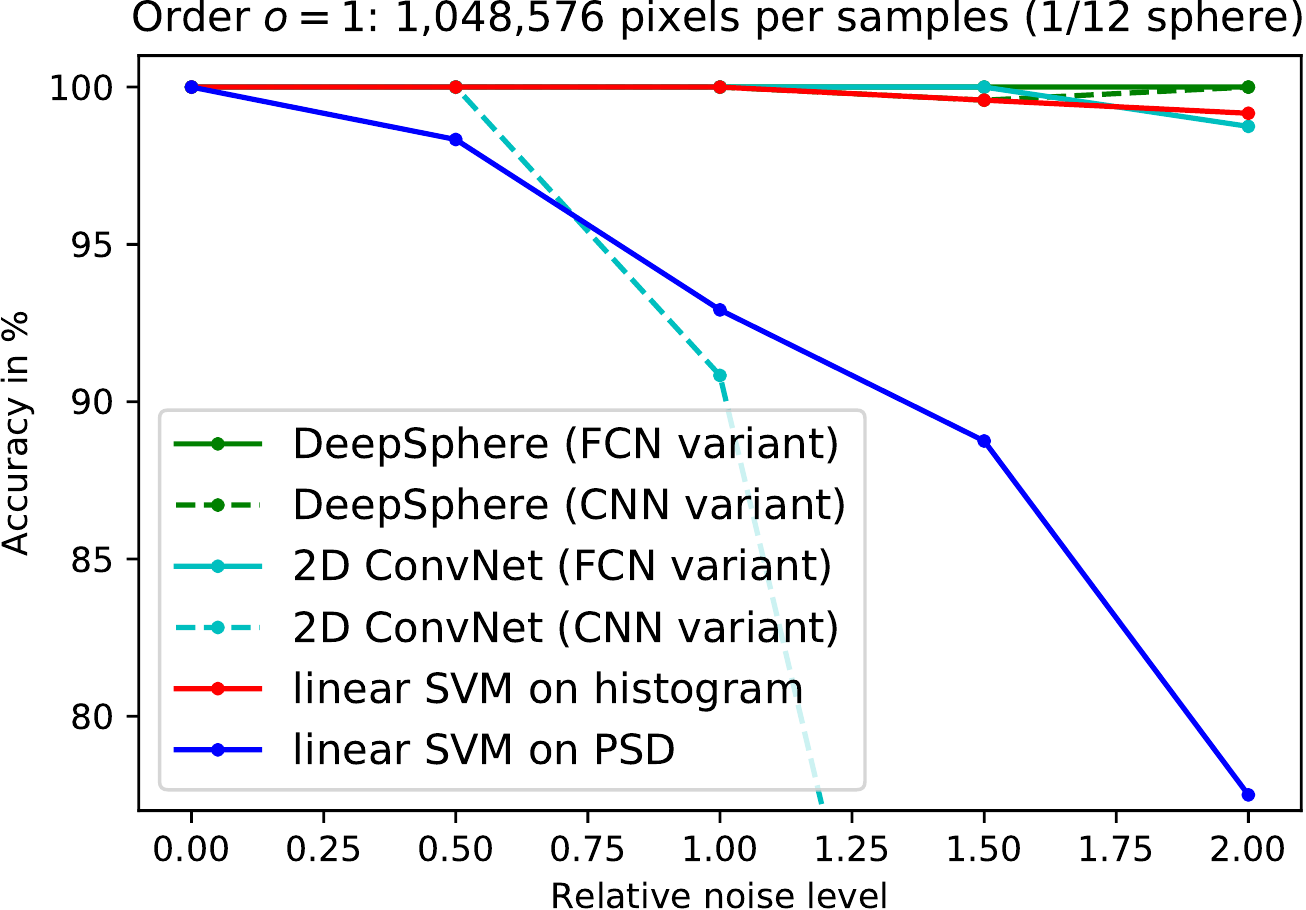}
	\hfill
	\includegraphics[width=0.32\linewidth]{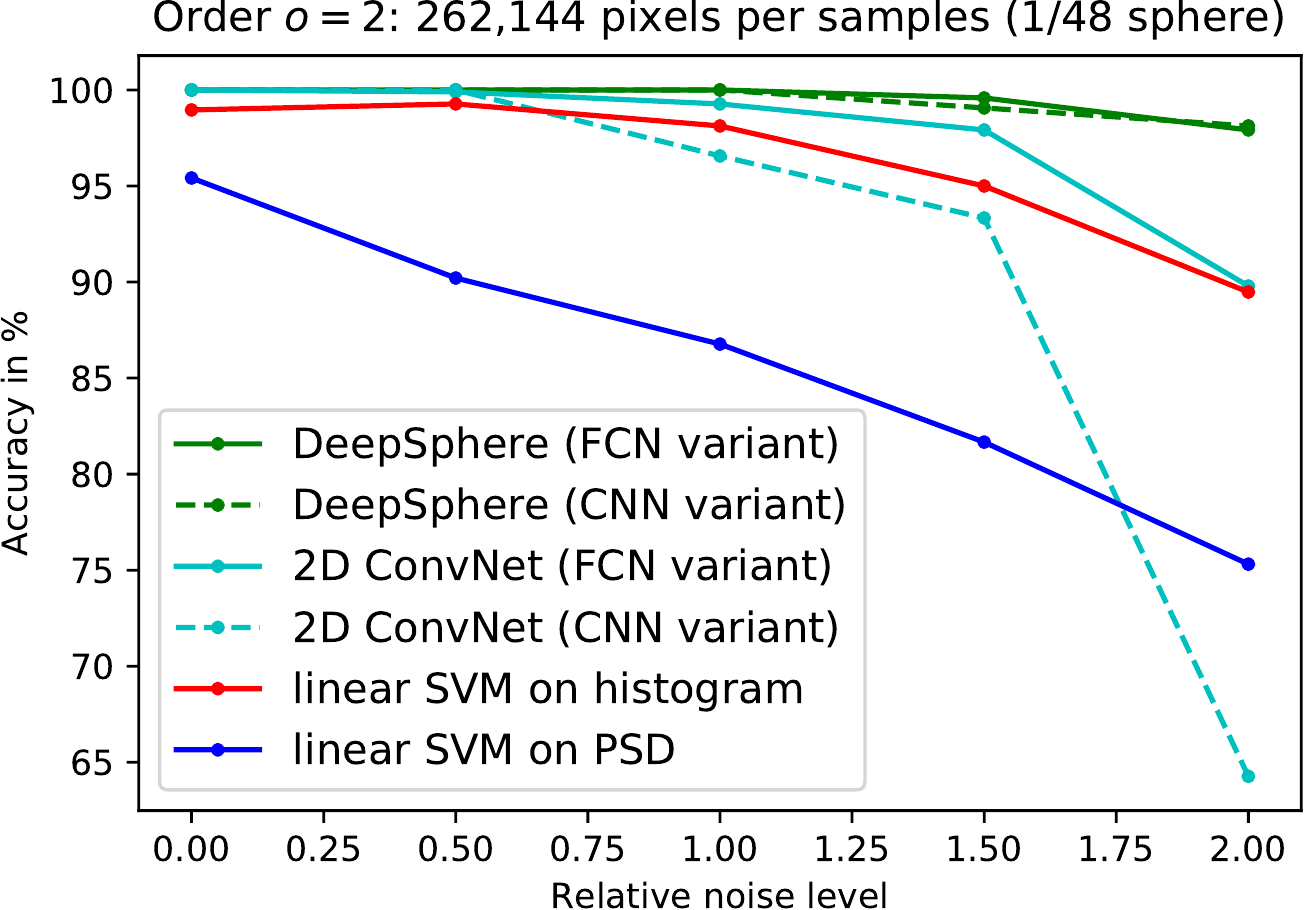}
	\hfill
	\includegraphics[width=0.32\linewidth]{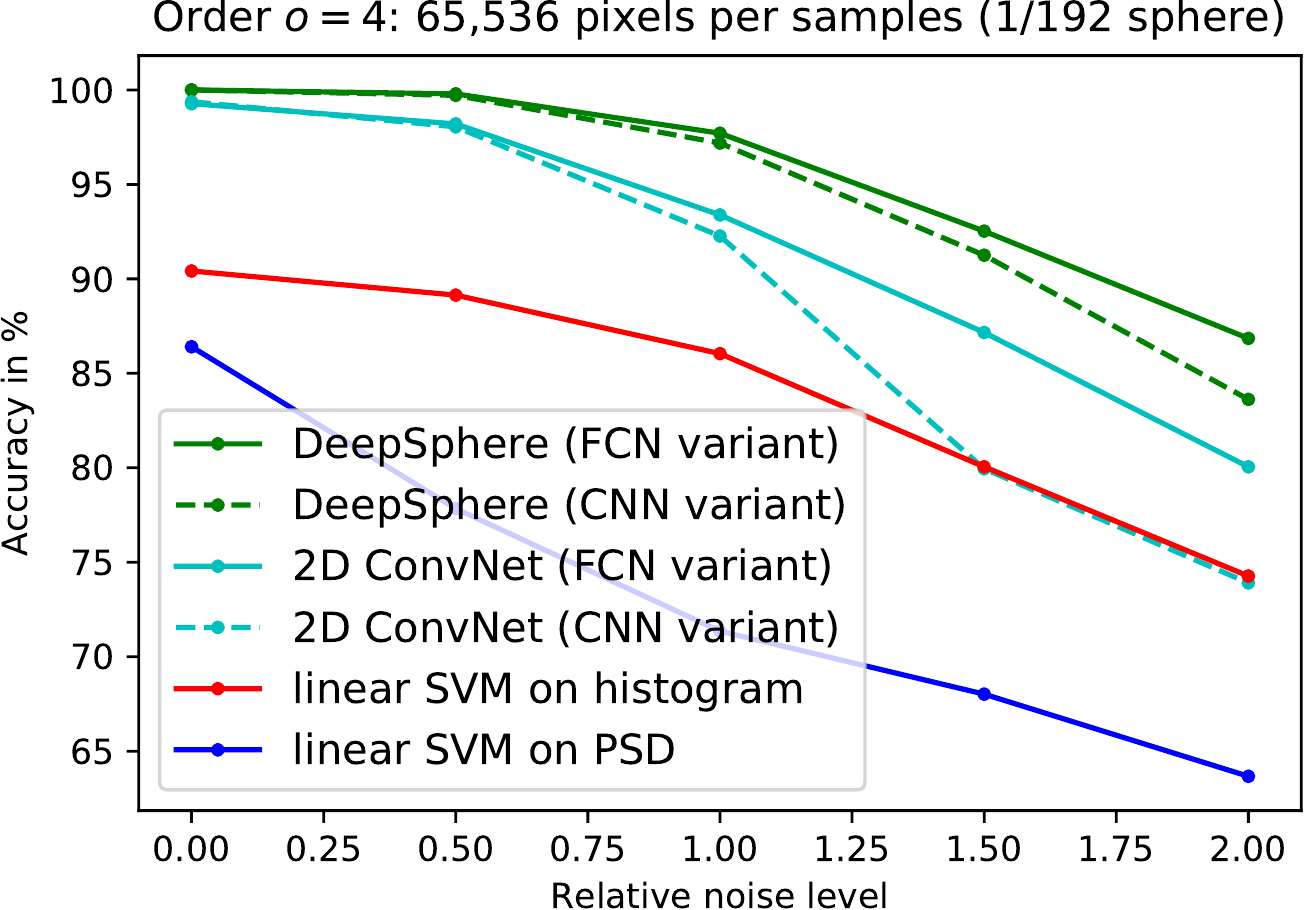}
	\caption{
		Classification accuracy of the fully convolutional variant of DeepSphere (DeepSphere FCN), the standard convolutional variant of DeepSphere (DeepSphere CNN), the fully convolutional 2D ConvNet (2D ConvNet FCN), the standard 2D ConvNet (2D ConvNet CNN), the support vector machine (SVM) with the power spectral density (PSD) as features, and the SVM with the histogram as features.
		The difficulty of the task depends on the level of noise and the size of a sample (that is, the number of pixels that constitute the sample to classify). Order $o=1$ corresponds to samples which area is $\frac{1}{12}=8.1\%$ of the sphere, order $o=2$ to $\frac{1}{12 \times 2^2} = 2.1\%$, and order $o=4$ to $\frac{1}{12 \times 4^2} = 0.5\%$. The standard deviation of the added Gaussian noise varies from zero to $2\times$ the standard deviation of pixel's values in the noiseless maps.
		It is clear from those results that the noise makes the problem harder, and that having more pixels available to classify a sample makes the problem easier (the classifier having more evidence to make a decision). The FCN variant of DeepSphere beats the CNN variant by being invariant to rotation. Both variants largely beat the 2D ConvNet and the two SVM baselines.
	}
	\label{fig:results}
\end{figure*}

\figref{results} compares the classification accuracy of DeepSphere (FCN and CNN variants), and the three baselines across five noise levels and three sample sizes.
We indeed observe that the problem is made more difficult as the sample size decreases, and the noise increases.
As fewer pixels are available to make a decision about a sample, the algorithms have access to less information and thus cannot classify as well.
As the noise increases, the useful information gets diluted in irrelevant data, i.e., the signal-to-noise ratio (SNR) diminishes.

As the cosmological parameters were chosen for the maps to have similar PSDs (see \figref{psd_sigma3}), it is reassuring to observe that an SVM with those as features has difficulties discriminating the models.
Other statistics are therefore needed to solve the problem. Using histograms as features gives a significant improvement.
Performance is very good for larger maps and deteriorates for the smaller cases with increased noise level, reaching 80\% at the highest noise level for $o=4$.
The histogram features contain information about the distribution of pixels, which clearly varies for the two classes.
They do not, however, include any spatial information.
DeepSphere (both the FCN and CNN variants) shows superior performance over all configurations.
The gap widens as the problem becomes more difficult.
This is likely due to DeepSphere learning features that are adapted to the problem instead of relying on hand-designed features.
The accuracy of DeepSphere is $>97\%$ for orders $o=1$ and $o=2$, for all noise levels, and starts to deteriorate for $o=4$, reaching 90\% for the highest noise level.

It may seem unfair to compare DeepSphere, who has access to raw pixels, with SVMs who only see limited features (histograms and PSDs).
We however trained an SVM on the raw pixels and were unable to obtain over $60\%$ accuracy in any of the three noiseless cases.

The 2D ConvNet fairs in-between the SVMs and DeepSphere.
Its relatively low performance probably come from the following drawbacks.
First, it does not exploit the rotational invariance of the problem.
We observed that the learned convolution kernels were not radial.
Second, the 2D projection distorts the geometry, and the NN has to learn to compensate for it, which comparatively requires more training data.

As expected, the FCN variants outperform the CNN variants (both for DeepSphere and the 2D ConvNet).
This may seem counterintuitive as the CNN is a generalization of the FCN and hence should be able to learn the same function and provide at least equivalently good results.
Nevertheless, as discussed in \secref{architecture}, the larger number of parameters incurred by the increased flexibility requires more training data to learn the rotation symmetry.
The superior performance of the FCN is an empirical validation that these maps are stationary with a small-support radial autocorrelation function, stemming from the fact that the mass distribution is homogeneous and isotropic.
It also implies that this classification problem is invariant to rotation.
Hence, a pixel can be statistically classified using only its surrounding, and those local predictions can be averaged to vote for a global consensus.
The CNN variant, however, may be better for data that does not have this property, as the architecture should always be adapted to the data and task (see \secref{architecture}).

While testing the FCN and CNN variants of DeepSphere, we made the following empirical observations.
First, training was more stable (in the sense that the loss decreased more smoothly) when using Chebyshev polynomials, as in \eqnref{graph_convolution_cheby}, rather than monomials, as in \eqnref{graph_convolution_monomial}.
Nevertheless, we could not observe a significant difference in accuracy after convergence.
Second, using $\ell_2$ regularization does not help either with performance or training of the models, mostly because the models are not over-fitting.
Third, we recommend initializing the Chebyshev coefficients with a centered Gaussian distribution with standard deviation $\sqrt{2/(F_{in} \times (K + 0.5))}$, where $K$ is the degree of the Chebyshev polynomials and $F_{in}$ the number of input channels.
This standard deviation has the property of keeping the energy of the signal more or less constant across layers.\footnote{We derived this rule from the principles presented in \cite{glorot2010understanding}, the Chebyshev polynomial equations, and some empirical experiments.}
Finally, we observe that scaling the Laplacian's eigenvalues between $[-a, a]$, where $0 < a \leq 1$, significantly helps in stabilizing the optimization. We use $a = 0.75$ in our experiments.

\begin{figure}
	\centering
	\includegraphics[width=0.85\linewidth]{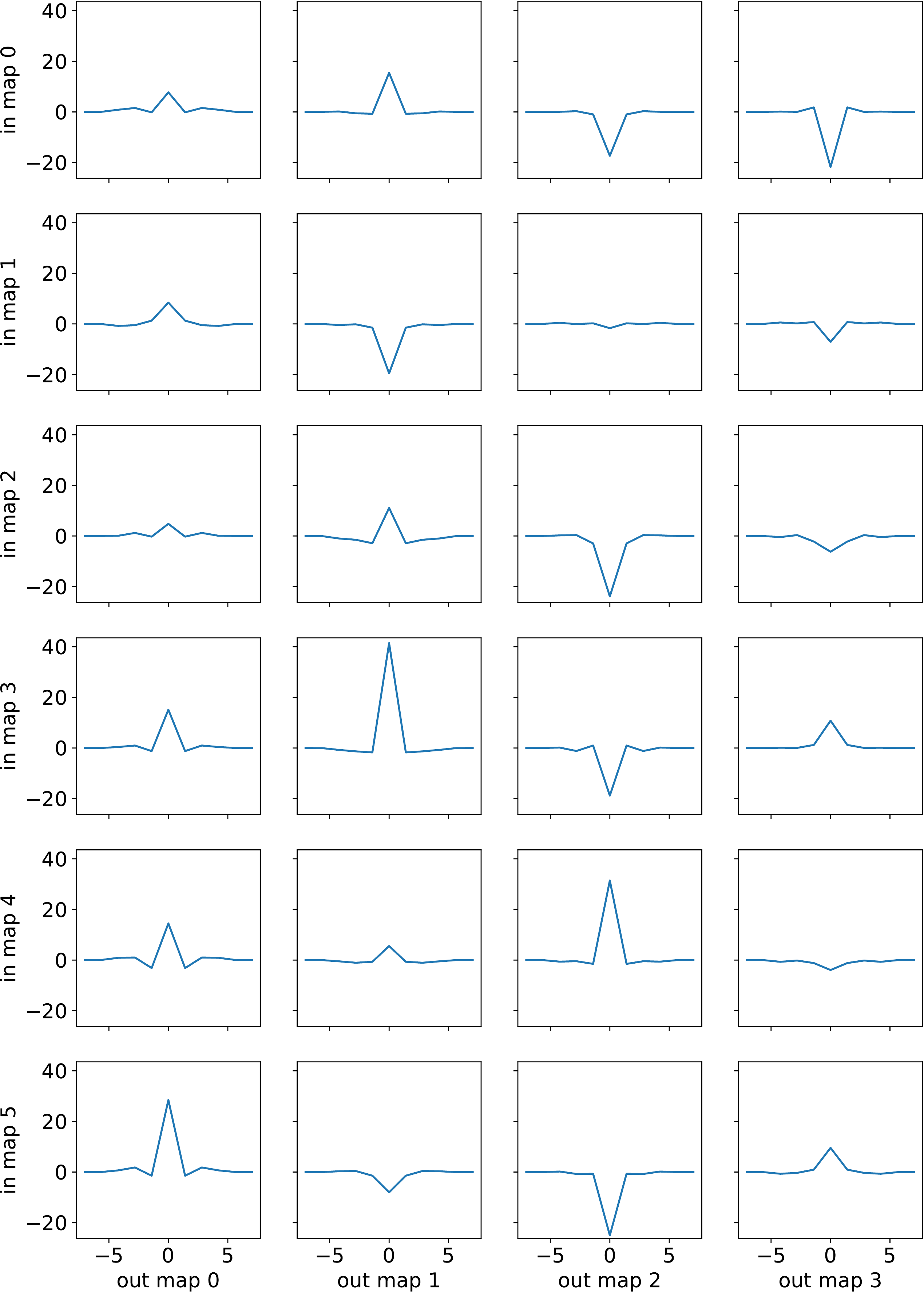}
	\vspace{0.5cm}\\
	\includegraphics[width=0.85\linewidth]{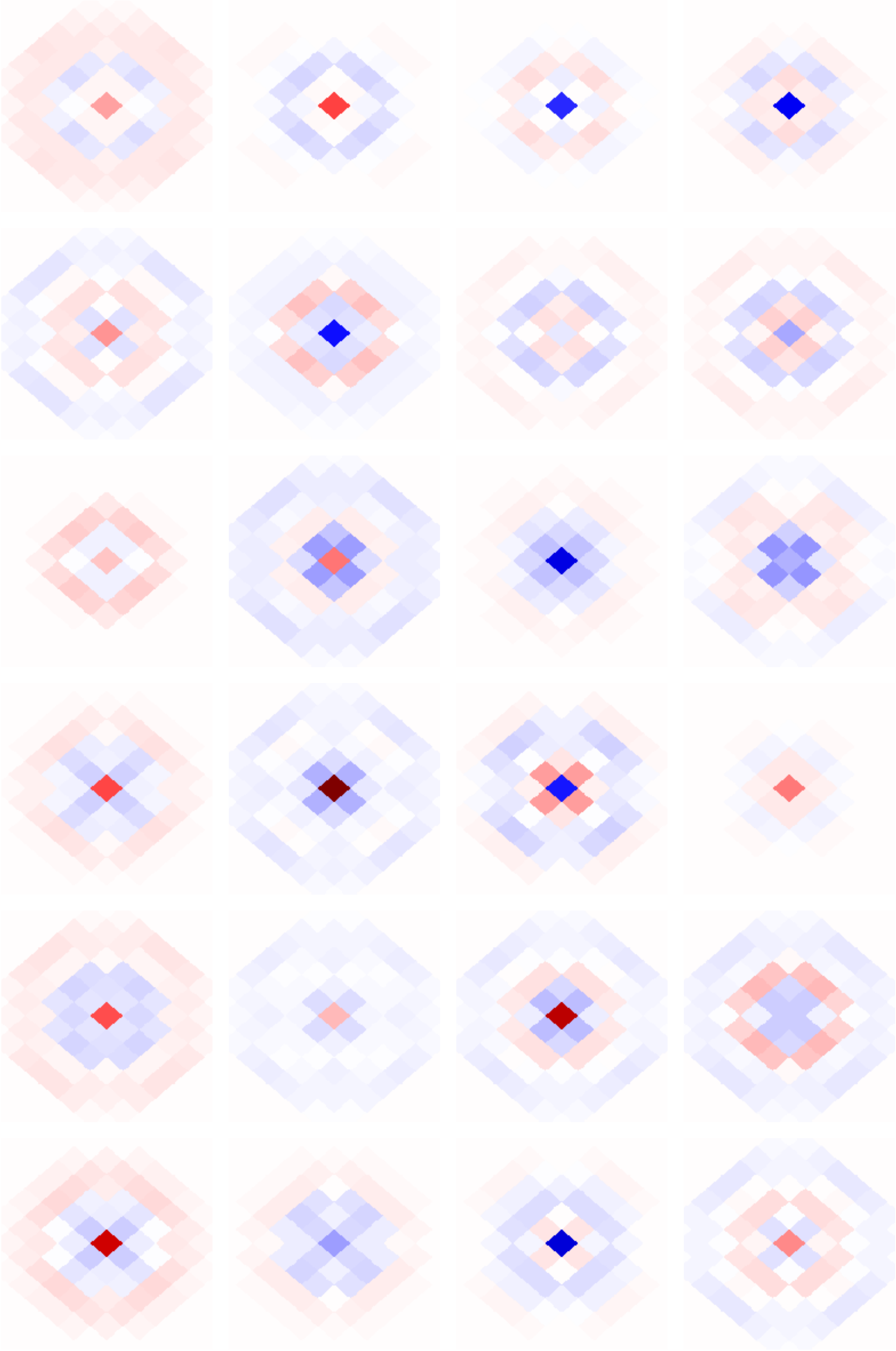}
	\caption{Random selection of 24 learned filters from the fifth spherical convolutional layer $L_5$. Top: section. Bottom: gnomonic projection.
	The structure in the filters resembles peaks, which is not unexpected, given that the convergence maps largely consist of concentrated clumps.}
	\label{fig:learned_filter}
\end{figure}

\subsection{Filter visualization}

A common visualization to introspect and try to understand how a CNN function is to look at the learned filters.
Since our construction leads to almost (up to sampling irregularities) spherical filters, we plot in \figref{learned_filter} both the radial profile and a gnomonic projection on a plane of a random selection of learned filters from the last layer of the network.
While those particular filters were obtained from the experiment with order $o=2$ and a relative noise level of $2$, all trained networks presented visually similar patterns.
Details of how the convolution kernels are plotted are described in \ref{sec:filter_visualization}.
While it is usually difficult to interpret the shape of the filters, especially given the type of data, we can notice that they often have a ``peak''-like structure.
An example of filter interpretation was demonstrated in \citep{Ribli2018learningfrom}.

\section{Conclusion}
\label{sec:conclusion}

We present DeepSphere, a convolutional neural network defined on the sphere with HEALPix sampling, designed for the analysis of cosmological data.
The main contributions of this paper are (i) to show that spherical CNNs are a great NN architecture for cosmological applications, and (ii) that a graph-based spherical CNN has certain undeniable advantages.
The core of our method is the use of a graph to represent the discretised sphere.
This allows us to leverage the advantages of the graph convolution.
It is both efficient, with a complexity of $\bO(N_{pix})$, and flexible, which allows DeepSphere to efficiently work on a partial sphere.
The spherical properties of the domain are well captured: the graph Fourier modes are close to the spherical harmonics.
Filters are restricted to be radial for the convolution operation to be equivariant to rotation.
DeepSphere can then be made either equivariant or invariant to rotation.
Equivariance is not perfect as small imprecisions due to the sampling cause the action of a graph filter to slightly depend on the location.
However, we do not expect that to cause problems for practical applications.

We demonstrate that DeepSphere systematically and significantly outperforms three benchmark methods on an example problem of cosmological model discrimination with weak lensing mass maps, designed similarly to \citep{schmelze2017cosmologicalmodel}.
The maps were produced from two cosmological models with varying $\sigma_8$ and $\Omega_m$, chosen to follow the typical weak lensing degeneracy in these parameters, so that the power spectra for these models are similar.
We compared the performance of DeepSphere versus that of a classical 2D CNN and two SVM classifiers (one trained on the spherical harmonics power spectrum, and the other on the pixel density histogram).
DeepSphere performs better than the three baselines for all considered cases.
The advantage is small for large, noise-free maps, and grows up to 10\% for smaller, noisier data.

Spherical CNNs are so far mostly used for omnidirectional imaging.
Many scientific fields, such as weather or climate modelling, would however benefit from an efficient and versatile spherical CNN.
With this work, we hope to demonstrate and help democratize the use of those tools for spherical data analysis.
We publish the code as a small and easy-to-use python package.
The code was designed so that DeepSphere can easily be used for typical machine learning tasks, such as classification and regression, both dense and global.

As future work, it would be interesting to further investigate the relation between the graph Fourier basis and the spherical harmonics.
The goal would be to find edge weights such that the graph Laplacian converges (or is equivalent) to the Laplace-Beltrami (up to a certain bandwidth).
Ideally, that should work for any sampling of the sphere.
That would enable a truly rotation equivariant graph convolution and high-precision filtering using the graph rather than the SHT (for speed).
Finally, a comparison of DeepSphere to other spherical CNN formulations, with different sampling schemes, would be worthwhile.

The fact that a graph representation of the discretised sphere enables an efficient convolution relaxes the iso-latitude constraint on the design of sampling schemes which aim to enable fast convolutions.
In the long term, graphs might enable researchers to consider sampling schemes with different trade-offs, or remove the need for schemes and interpolation altogether and simply consider the positions at which measures were taken.

\section*{Acknowledgements}

This work was supported by a grant from the Swiss Data Science Center (SDCS) under project \textit{DLOC:  Deep Learning for Observational Cosmology} and grant number 200021\_169130 from the Swiss National Science Foundation (SNSF).
We thank Jean-François Cardoso for helpful discussions about the spherical harmonics.
We thank Alexandre Refregier, Adam Amara, Thomas Hofmann, and Fernando Perez-Cruz for advice and helpful discussions.
We thank Hamsa Padmanabhan for testing use cases of the code.
Finally, we thank the two anonymous reviewers who provided extensive feedback that greatly improved the quality of this paper.

\appendix

\section{Graph Fourier modes and spherical harmonics}
\label{sec:comparison_spherical_harmonics}

\begin{figure}[t!]
	\centering
	\includegraphics[width=\linewidth]{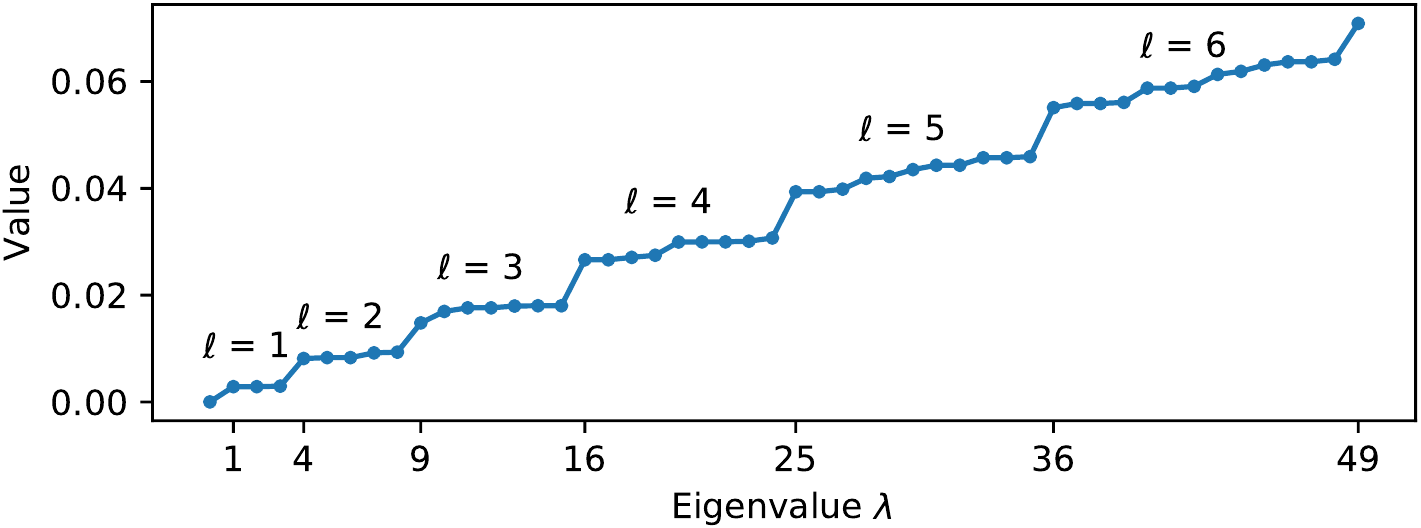}
	\caption{The eigenvalues $\bLambda$ of the graph Laplacian $\L = \U \bLambda \U\trans$, which corresponds to squared frequencies, are clearly organized in groups. Each group corresponds to a degree $\ell$ of the spherical harmonics. Each degree has $2\ell + 1$ orders. See also \figref{graph_harmonics}.}
	\label{fig:graph_eigenvalues}
\end{figure}

\begin{figure}[t!]
	\centering
	\includegraphics[width=\linewidth]{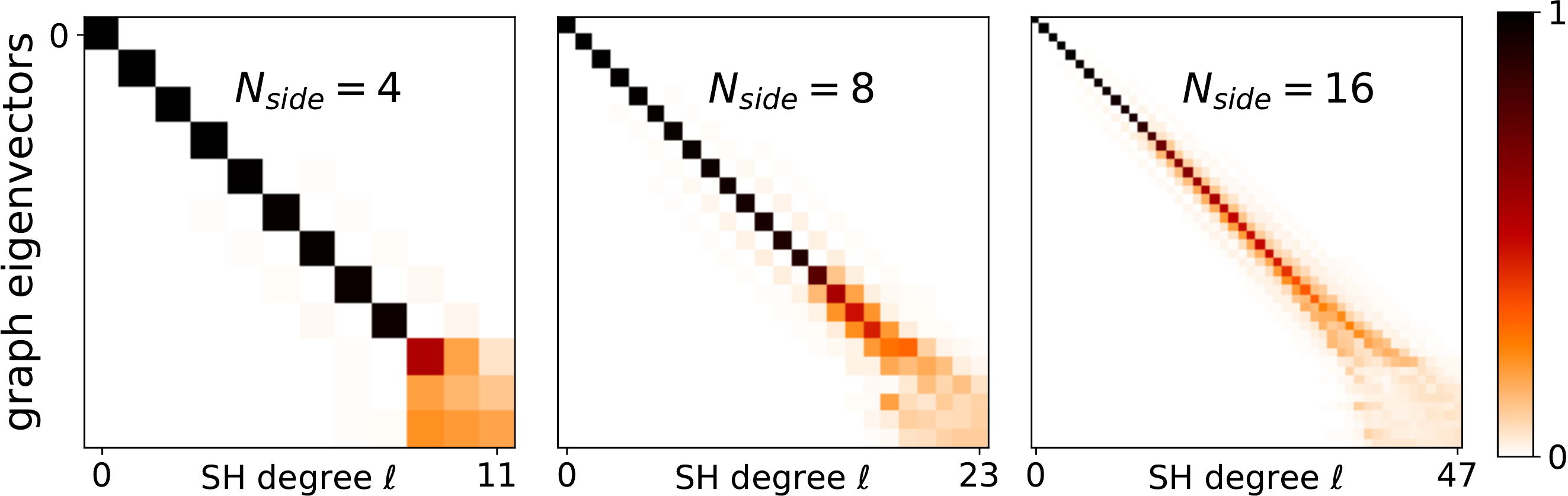}
	\caption{Correspondence between the subspaces spanned by the graph Fourier modes and the spherical harmonics.
		First, we compute the power spectral density (PSD) of each graph eigenvector with the SHT.
		Second, as there is $2\ell+1$ spherical harmonics for each degree $\ell$, we sum the contributions of the corresponding $2\ell+1$ graph eigenvectors.
		The matrix shows how the subspaces align: the Fourier modes span the same subspaces as the spherical harmonics in the low frequencies, and the Fourier modes leak towards adjacent frequency bands at higher frequencies.
		While there is a systematic error, the Fourier modes align at higher frequencies as $N_{side}$ increases.}
		\label{fig:subspace_harmonics_eigenvectors}
\end{figure}

The first 16 eigenvectors $[\b u_1, \ldots, \b u_{16}]$ of the graph Laplacian $\L$, forming the lower part of the graph Fourier basis $\U$, are shown in \figref{graph_harmonics}.
Let us further observe the spectral properties of our constructed spherical graph laplacian $\L$.
Its eigenvalues, shown in \figref{graph_eigenvalues}, are clearly organized in frequency groups of $2\ell + 1$ orders for each degree $\ell$.
We remind the reader that the amplitude of the Laplacian eigenvalue is proportional to the sum of the variations of its associated eigenvector. Furthermore, all spherical harmonics with the same order $\ell$ have the same variation. Hence, the fact that the Laplacian eigenvalues are grouped in blocks of size $2\ell + 1$ is a hint that the graph eigenvectors approximate the spherical harmonics.
In order to push the comparison one step further, we show the correspondence between the subspaces spanned by the graph Fourier modes and the spherical harmonics in \figref{subspace_harmonics_eigenvectors}.

While these indications show that the constructed graph Fourier basis approximates well the spherical harmonics, one should not forget that the small irregularities in the sampling (non-constant number of neighboring pixels and varying distance between pixels, see \figref{healpix_graph_4}) have an important effect on the graph Fourier modes.
First, we believe that they are responsible for energy leaking across frequency bands in \figref{subspace_harmonics_eigenvectors}.
Second, as the resolution is increased with $N_{side} \rightarrow \infty$ and $N_{pix} \rightarrow \infty$, we are still unsure if the eigenvectors would converge towards the spherical harmonics.
The theoretical study of those phenomenons is left as future work.
Third, counter-intuitively, some eigenvectors will be localized~\citep{perraudin2018global}, i.e., they will span a small part of the sphere.
Those discrepancies result in a convolution operation that is not exactly equivariant to rotation.
Nevertheless, our experiments suggest that these downsides do not have an important effect on the convolution nor hinder the performance of the NN.

\section{Example: heat diffusion}
\label{sec:heat_diffusion}
\label{sec:filter_visualization}

\begin{figure}[ht!]
  \centering
  \includegraphics[width=\linewidth]{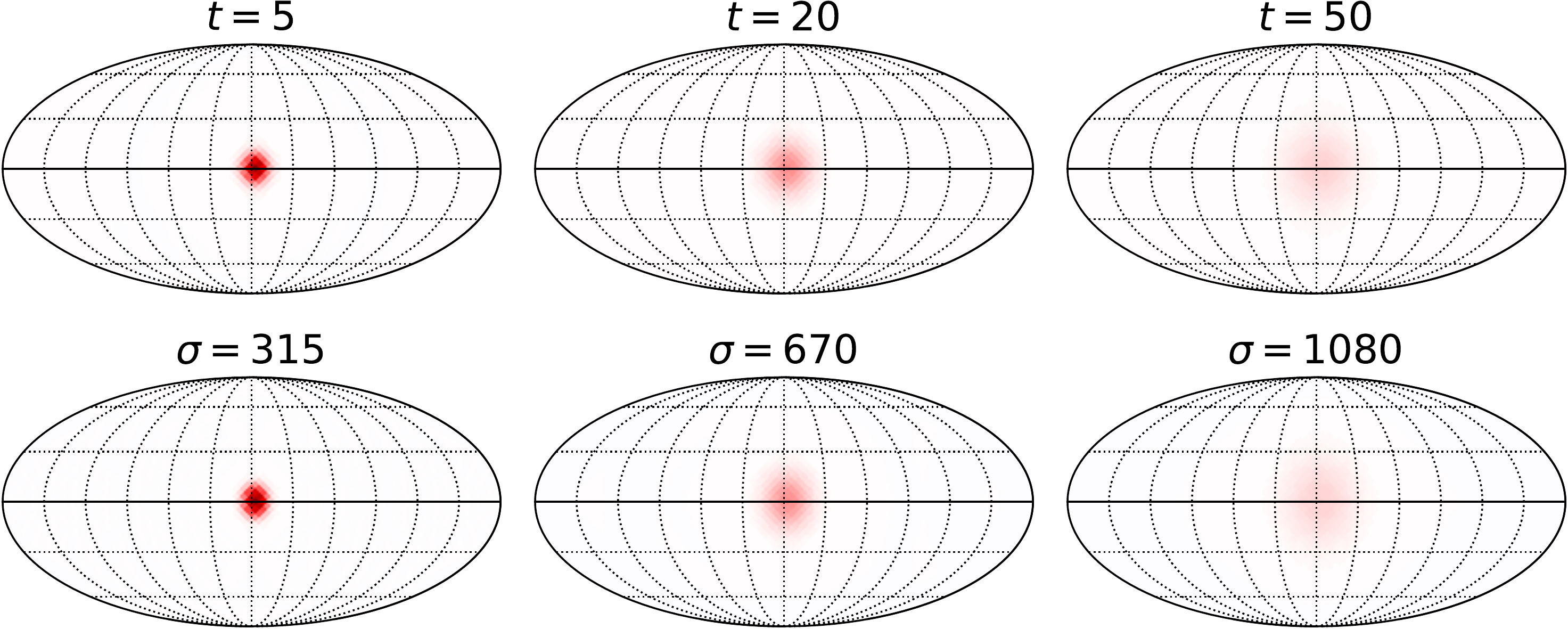}
  \caption{Comparison of convolution with the graph and the spherical harmonics ($N_{side} = 16$).
  Top: Diffusion of a unit of heat for different times $t$ using the graph.
  Bottom: spherical symmetric Gaussian smoothing for different $\sigma$ (arcmin).
  Relative difference between graph convolution and spherical smoothing: $10.4$\%, $4.8$\%, $3.8$\%.
}
  \label{fig:gaussian_filters_comparizon}
\end{figure}

\begin{figure}[t]
	\centering
	\includegraphics[width=\linewidth]{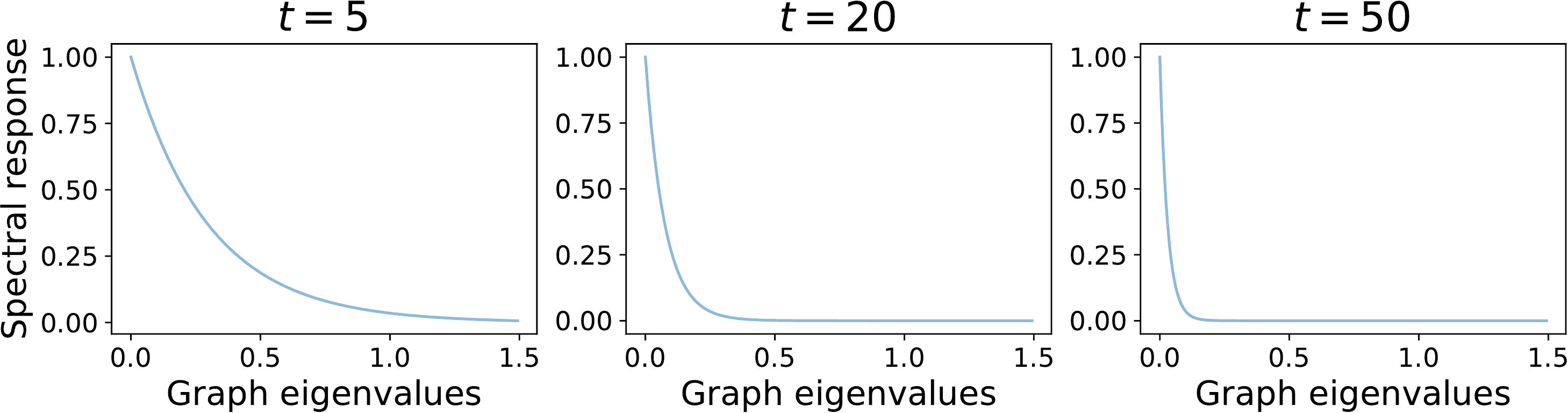} \\
	\vspace{0.2in}
	\includegraphics[width=\linewidth]{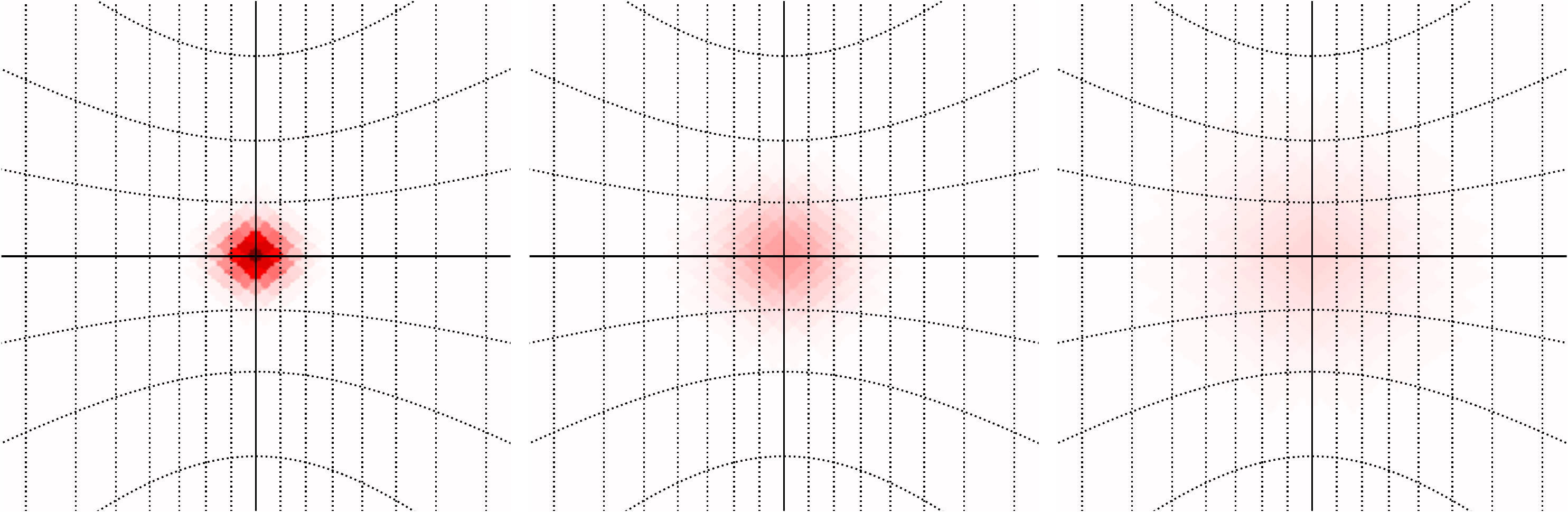} \\
	\vspace{0.2in}
	\includegraphics[width=\linewidth]{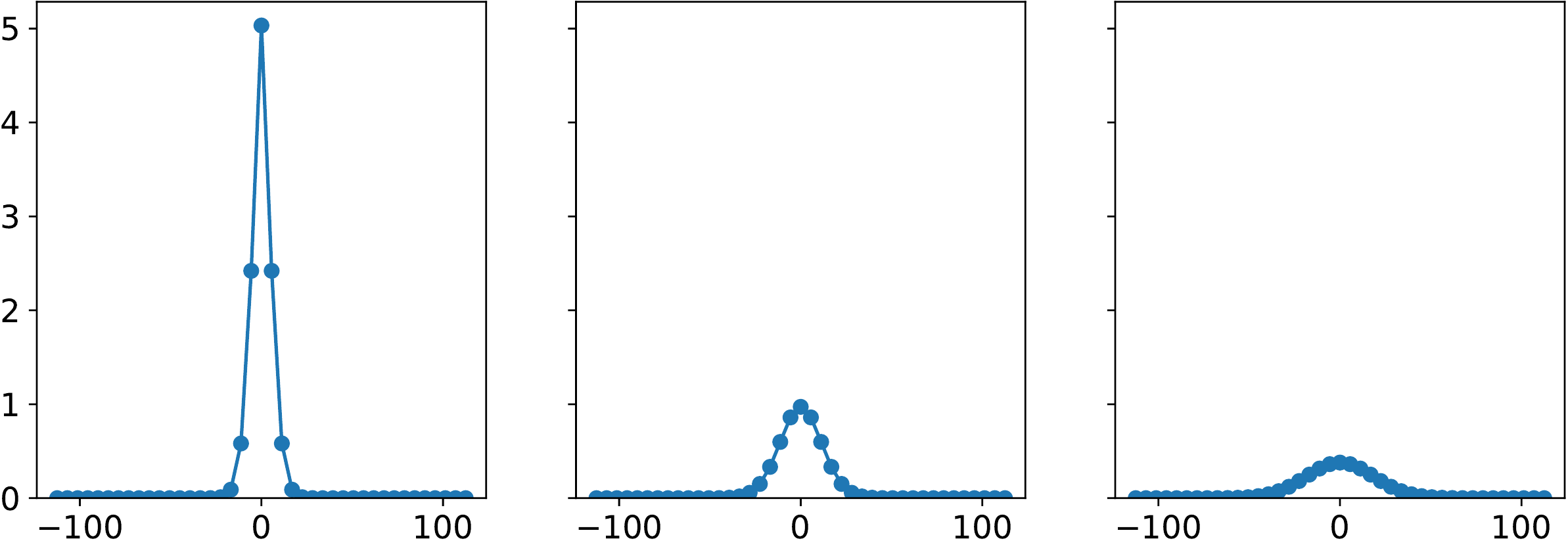}
	\caption{Visualization of the convolution kernel $K_t(x)=e^{-\tau t x}$. Top: spectral domain. Middle: gnomonic projection. Bottom: cross-section along the equator.}
	\label{fig:gaussian_filters_visualization}
\end{figure}

Let us consider the heat diffusion problem
\begin{equation}
  \tau \L \b{f}(t) = - \partial_t \b{f}(t),
  \label{eqn:heat_equation}
\end{equation}
where $\b{f}: \R_+ \rightarrow \R^{N_{pix}}$. Given the initial condition
$\b{f}(0)$, the solution of~\eqnref{heat_equation} can be expressed as
\begin{equation*}
  \b{f}(t) = e^{-\L \tau t} \b{f}(0) = \U e^{-\bLambda \tau t} \U\trans \g{f}(0) = K_t(\L) \b{f}(0),
\end{equation*}
which is, by definition, the convolution of the signal $\b{f}(0)$ with the kernel $K_t(\lambda)=e^{-\tau t \lambda}$. Since the kernel $K_t$ is applied to the graph eigenvalues $\bLambda$, which can be interpreted as squared frequencies, it can also be considered as a generalization of the Gaussian kernel on the sphere. \figref{gaussian_filters_comparizon} shows the effect of the convolution by diffusing a unit of heat for $\tau=1$ at various times $t$. By comparing the graph convolution with the spherical symmetric Gaussian smoothing, we observe that both techniques lead to similar results (see \figref{gaussian_filters_comparizon}). While the graph convolution remains different from the spherical convolution, this small experiment shows that, providing the correct parameters, the graph convolution can approximate the spherical convolution.
\figref{gaussian_filters_visualization} should give more insights about the graph kernel $K_t$.

Furthermore, the harmonic resemblance is another sign that the constructed graph is able to capture the spherical structure of the HEALPix sampling.
In some applications, where the exactitude of the convolution is not a requirement, such as de-noising, graph convolution could be used instead of using the spherical harmonics.
In a neural network setting, the fact that the graph convolution is not exactly equivariant to rotation is probably compensated by the fact that convolution kernels are learned.

We further leverage this example to present three visualizations of a convolution kernel.
The first visualization (\figref{gaussian_filters_visualization}, top) is a plot of the kernel function, i.e., it shows $K_t$ evaluated at the graph eigenvalues $\text{diag}(\bLambda)$.
Note that small eigenvalues $\lambda$ correspond to small frequencies, i.e., to spectral modes of low variation.
The second visualization (\figref{gaussian_filters_visualization}, middle) shows the kernel localized on the sphere (gnomonic projection).
As the kernel $K_t$ is defined in the graph spectral domain, it was convolved with a Kronecker $\delta$ to obtain a representation on the sphere (see \secref{graph_convolution}).
The third visualization (\figref{gaussian_filters_visualization}, bottom) plots the section of the filter.
This visualization might be easier to read as the filters are isotropic (up to the irregularities of the sampling).
The plot is obtained by convolving a Kronecker $\delta$ on the vertices along the equator.
Note that, because of the small irregularities in the HEALPix sampling, the second and third methods are subject to small variations depending on the chosen position of the Kronecker $\delta$.

\section{Border effects}
\label{sec:border_effects}

\begin{figure}[ht!]
	\centering
	\includegraphics[width=\linewidth]{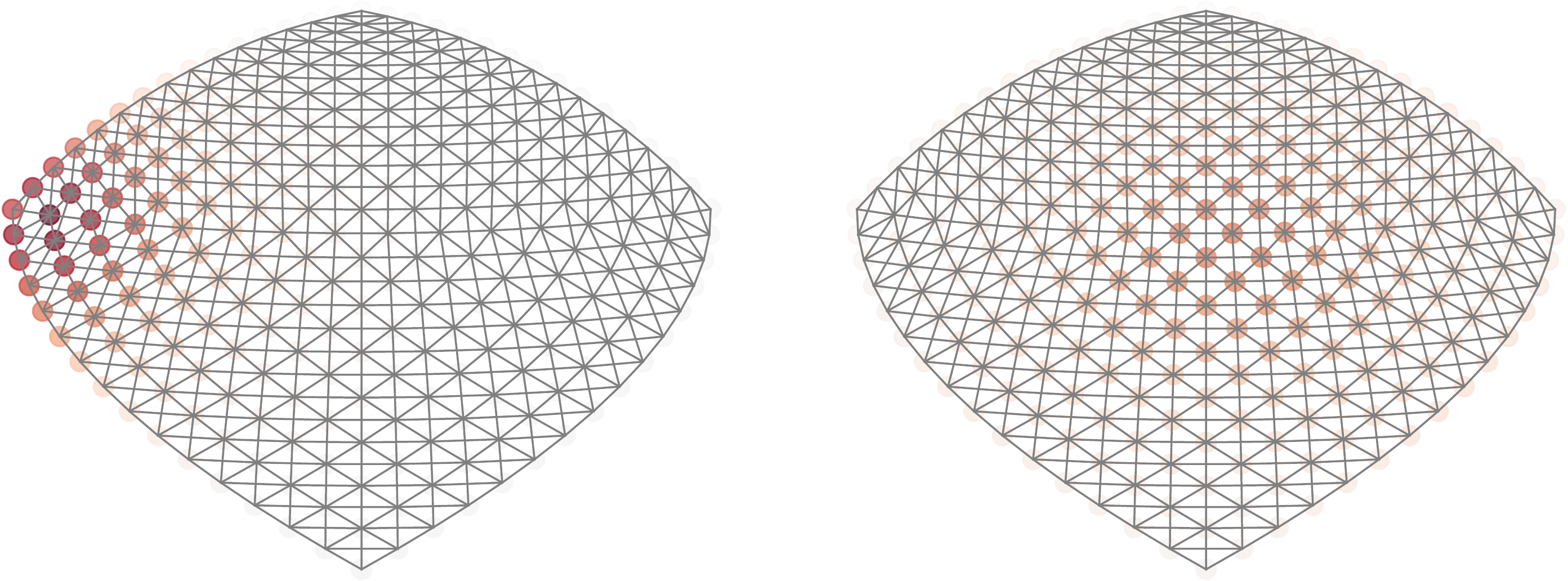}
	\caption{Convolution kernel (also called filter) localized in the center and left corner of a graph built from $1/12^\text{th}$ of the sphere at $N_{side} = 16$.
		A filter $h$ is localized on a pixel $i$ as $\T_i h = h(\L) \b \delta_i$ (see \secref{graph_convolution}, equation~\eqnref{graph_convolution_spatial}).
	The filter is not isotropic anymore when localized on the corner as the graph representation of a manifold assumes reflective border conditions.}
	\label{fig:border_effects}
\end{figure}

The graph setting used throughout this contribution corresponds to assumed reflective border conditions.
While that is irrelevant when working on the complete sphere (as it has no border), it slightly affects the convolution operation when only a part of the sphere is considered.
As depicted in \figref{border_effects}, a filter localized near a border (via $h(\L) \b \delta_i$) is no longer isotropic.
These border effects can, however, be mitigated by padding with zeros a small area around the part of interest (in which case they become similar to border effects in classical CNNs).
We however do not expect these effects to cause any problem as long as the data samples cover the same area.

\section{Projection onto a 2D plane}
\label{sec:projection}

In order to compare the DeepSphere architecture with a 2D CNN, we needed to project the spherical maps onto a plane.
A radial projection is computationally so expensive that, because of data augmentation, it is significantly more expensive than the actual training of the NN.
Hence, we decided to leverage the geometrical properties of HEALPix to create another projection.
Remember that HEALPix models the sphere as $12$ pixels that are then subdivided into $4$ recursively.
Eventually, each of these $12$ bases resembles a square grid.
See \figref{border_effects} to visualize it.
As our classification problem involves a twelfth of the sphere at most, we can use this natural structure to simply assign each HEALPix pixel to a pixel of a 2D grid of size $N_{sides} \times N_{sides}$.
This operation is computationally cheap and probably spreads the projection error relatively evenly across the resulting image.
This projection has actually been independently proposed (while this paper was under review) for a spherical CNN on HEALPix for cosmology \cite{krachmalnicoff2019convolutional}.

\section{Augmentation of the dataset for the SVM classifier}
\label{sec:dataset_augmentation}

For the trained network to be robust to noise, we employ the following data augmentation technique: a random realization of Gaussian noise is added to each sample before it is fed to the NN.
Besides robustness to noise, the network is less likely to over-fit the training set as it never sees the same sample twice.

\begin{figure}[ht!]
	\centering
	\includegraphics[width=\linewidth]{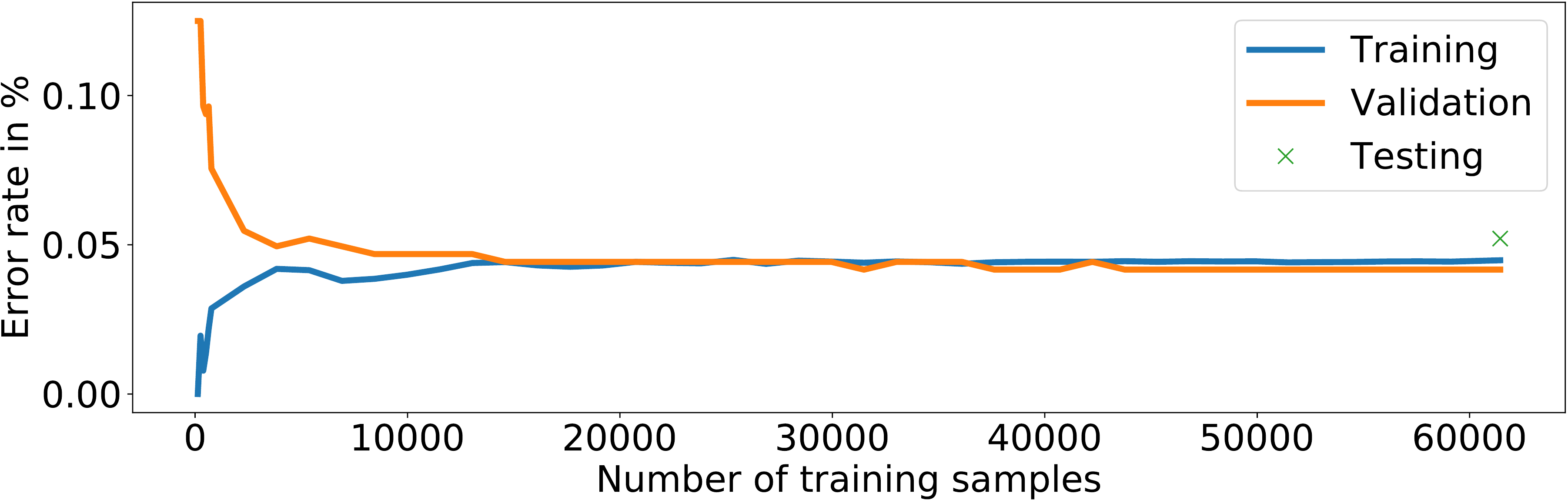}
	\caption{Error w.r.t.\ the number of training samples for histogram features and the linear SVM (setting: order $o=2$, relative noise level 1.5). The error clearly converged such that adding more samples will not improve the classifier. The validation error might be slightly below the training error as it was used to select the hyper-parameters.}
	\label{fig:hist_error_evolution}
\end{figure}

For a fair comparison between DeepSphere and the baselines, we have to ensure that the SVM classifier has access to the same amount of training data as DeepSphere. That is potentially an infinite number of samples. Hence, we fit different classifiers using various training set sizes until we experimentally observe that increasing the amount of training data does not improve performance. An example convergence of the training and validation errors is shown in \figref{hist_error_evolution}.

\section{Convergence Mass Maps}
\label{sec:convergence_mass_maps}

In the following we describe the weak lensing map-making pipeline \pkg{UFalcon} (ultra fast lightcone) developed in \citep{sgier2018fastgeneration} and the way it was used to generate data for this work.
The data consist of two-dimensional mass maps on the full sphere, which represent the dimensionless, weighted and projected mass along the light of sight.
The matter content of the universe consists mostly of dark matter.
This kind of projected mass maps are typically measured using the weak gravitational lensing technique.
Gravitational lensing occurs when images of distant galaxies are distorted by the matter clumps between the galaxy and the observer, which act as lenses.
Measurements of the small, spatially coherent distortions of galaxy images can be used to infer the matter density between the observer and the background galaxies.
\citep[see][for a review of gravitational lensing]{bartelman2010gravitationallensing}.

In order to obtain a convergence map characterising the gravitational lensing of distance sources, one has to start with simulations of the dark matter distribution.
This is achieved through cosmological N-Body simulations, which evolve a chosen number of dark matter particles $N_\mathrm{part}^\mathrm{sim}$ within a specific simulation volume $V_\mathrm{sim}$ under the influence of gravity across cosmic time-scales. The simulation outputs the positions of the particles at user-specific time-steps. These time-steps correspond to redshifts $z$, by which the spectrum is shifted the observer would measure. A redshift of $z=0$ correspond to the present time and higher redshifts to earlier times.

\begin{figure}[ht!]
	\centering
	\includegraphics[width=\linewidth]{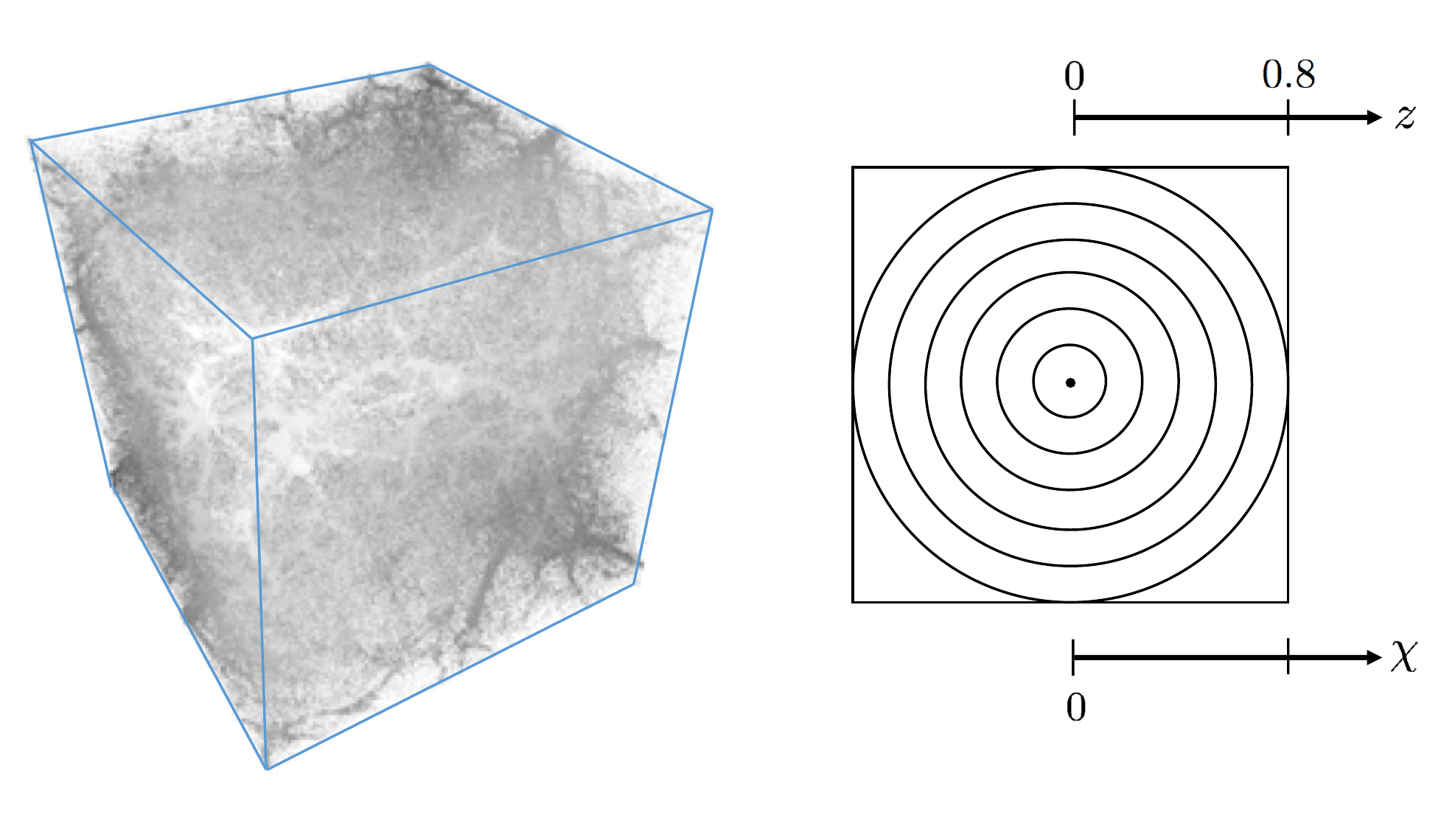}
	\caption{Left: Dark matter density field simulated using L-PICOLA for specific cosmological parameters. Right: Sketch of the lightcone constructed by arranging particles at different redshifts in concentric shells around the observer located at $z=0$.}
	\label{fig:sketch_lightcone_simulation}
\end{figure}

Since the convergence map characterises the dark matter lenses at different redshifts, one has to first construct a lightcone. The lightcone represents the volume of the universe around the observer, where the radius represents the distance the light from distant galaxies at different times and locations has to travel to the observer. Therefore, the particles outputted at different redshifts (time-steps) of the simulation are concentrically arranged in shells of constant redshift around the observer (each redshift $z$ corresponds to a comoving distance away from the observer $\chi(z)$), which is located at redshift $z=0$ (see \figref{sketch_lightcone_simulation}). The shells are then weighted by a weak lensing specific weight $W_b$ and projected onto the sphere for each pixel $\theta_\mathrm{pix}$, resulting in a convergence map
\begin{equation}
\kappa (\theta_\mathrm{pix}) \sim \sum_b W_b \frac{V_\mathrm{sim}}{N_\mathrm{part}^\mathrm{sim}} \frac{n_p (\theta_\mathrm{pix}, \Delta \chi_b)}{\mathcal{D}^2 (z_b)}  \, ,
\end{equation}
where $n_p$ represents the number of particles in a specific pixel on the shell $\Delta \chi_b$. The convergence map depends on the cosmological parameters.

The data generated for the present analysis is based on the fast approximate N-Body simulation code \pkg{L-PICOLA} \citep{howlett2015lpicola}. For our purposes, we chose a simulation volume $V_\mathrm{sim} = (4200 \, h^{-1}\, \mathrm{Mpc})^3$, a total number of particles $N_\mathrm{part}^\mathrm{sim} = 1024^3$ placed on a mesh of size $N_\mathrm{mesh}^\mathrm{sim} = 2048^3$. The simulation was then performed 30 times with different seeds for each of the two $\Lambda$CDM cosmological models described in section \secref{data} covering a redshift range from $z = 0.1$ to $z = 0.8$. The \pkg{L-PICOLA} were performed in lightcone mode, where the simulation output is already arranged as a lightcone. Here we chose a redshift-shell thickness of $\Delta z = 0.01$.

\section*{Bibliography}
\bibliographystyle{elsarticle-num}
\biboptions{sort}
\bibliography{biblio}

\end{document}